\documentclass[usegraphicx,usenatbib]{mn2e}

\begin{document}

\title[Cosmological parameters from CMB and 2dFGRS]
{
Cosmological parameters from CMB measurements 
and the final 2dFGRS power spectrum
}
\author[Ariel G. S\'{a}nchez et al.]
{\parbox[t]{\textwidth}{
Ariel G. S\'{a}nchez$^{1}$\thanks{E-mail: arielsan@oac.uncor.edu},
C. M. Baugh$^{2}$, 
W. J. Percival$^{3}$,
J. A. Peacock$^{3}$,
N. D. Padilla$^{4}$,  
S. Cole$^{2}$,
C. S. Frenk$^{2}$,
P. Norberg$^{5}$.}
\vspace*{6pt} \\ 
$^{1}$ Grupo de Investigaciones en Astronom\'{\i}a Te\'{o}rica y Experimental
(IATE), OAC, UNC, Argentina.\\
$^{2}$ The Institute for Computational Cosmology, 
Department of Physics, University of Durham, South Road, Durham DH1
3LE, UK.\\
$^{3}$ Institute for Astronomy, University of Edinburgh, Royal Observatory, 
Blackford Hill, Edinburgh EH9 3HJ, UK. \\
$^{4}$ Departamento de Astronom\'{\i}a y Astrof\'{\i}sica, Pontificia 
Universidad Cat\'olica, Vicuna Mackenna 4860 Santiago 22, Chile. \\
$^{5}$ Institut fur Astronomie, Department Physik, ETH Zurich, HPF G3.1, 
CH-8903 Zurich, Switzerland.\\
}
\date{Submitted to MNRAS}
\maketitle

\begin{abstract}
We derive constraints on cosmological parameters using the 
power spectrum of galaxy clustering measured from the final 
two-degree field galaxy redshift survey (2dFGRS) and a compilation of 
measurements of the temperature power spectrum and 
temperature-polarization cross-correlation of the cosmic microwave background 
radiation.  
We analyse a range of parameter sets and priors, allowing for  
massive neutrinos, curvature, tensors and general dark energy models. 
In all cases, the 
combination of datasets
tightens the constraints, with the most dramatic 
improvements found for the density 
of dark matter and the energy-density of dark energy.  
If we assume a flat universe, we find a matter density parameter of 
$\Omega_{\rm m}=0.237 \pm 0.020$, a baryon density parameter of 
$\Omega_{\rm b} = 0.041 \pm 0.002$, a Hubble constant of 
$H_{0}=74\pm2 \; {\rm kms}^{-1}{\rm Mpc}^{-1}$, a linear theory matter 
fluctuation amplitude of $\sigma_{8}=0.77\pm0.05$ and a scalar 
spectral index of $n_{\rm s}=0.954 \pm 0.023$ (all errors show 
the 68\% interval). 
Our estimate of $n_{\rm s}$ is only marginally consistent with the scale
invariant value $n_{\rm s}=1$; this spectrum is formally excluded at the
$95\%$ confidence level.
However, the detection of a tilt in the spectrum is sensitive to 
the choice of parameter space. 
If we allow the equation of state of the dark energy to float, 
we find $w_{\rm DE}= -0.85_{-0.17}^{+0.18}$, consistent with 
a cosmological constant. 
We also place new limits on the mass fraction of massive neutrinos: 
$f_{\nu} < 0.105$ at the 95\% level, corresponding  
to $\sum m_{\nu} < 1.2$~eV. 
\end{abstract}


\section{Introduction}

Since the turn of the millennium, we have witnessed a dramatic 
improvement in the resolution and accuracy of measurements of 
fluctuations in the temperature of the cosmic microwave background 
radiation (CMB). The discovery of features in the power spectrum of 
the CMB temperature, the acoustic peaks, marked the start of a new, 
data-rich era in cosmology (de Bernardis et~al. 2000; 
Hanany et~al. 2000). The relative positions and heights of the acoustic 
peaks encode information about the values of the fundamental 
cosmological parameters, such as the curvature of the universe or 
the physical density in cold dark matter and baryons.
Perhaps the most striking example of the progress achieved is the 
first year data from the WMAP satellite (Bennett et~al. 2003; 
Hinshaw et~al. 2003). 

The CMB data alone, however, do not constrain all of the 
fundamental cosmological parameters to high precision. 
Degeneracies exist between certain combinations of parameters which  
lead to indistinguishable temperature fluctuation spectra 
(Efstathiou \& Bond 1999). Some of these degeneracies 
can be broken by comparing theoretical models to 
a combination of the CMB data and  
other datasets, such as the power spectrum of galaxy clustering. 
At the same time as the new measurements of the CMB were obtained,  
two groundbreaking surveys of galaxies in the local 
Universe were being conducted. The two-degree field galaxy redshift 
survey (2dFGRS; Colless et~al. 2001; 2003) and the Sloan Digital Sky 
Survey (SDSS; York et~al. 2000; Abazajian et~al. 2005) are substantially larger than previous 
redshift surveys and allow the clustering of galaxies to be measured 
accurately on all scales. On large scales, the connection to 
theoretical models is most straightforward. 

Percival et~al. (2001) used the power spectrum of galaxy clustering 
measured from the 2dFGRS to constrain the ratio of the baryon to matter 
density, $\Omega_{\rm b}/\Omega_{\rm m}$, and the matter density, 
$\Omega_{\rm m}h$. 
Efstathiou et~al. (2002) used a compilation of pre-WMAP CMB data and 
the Percival et~al. measurement of the galaxy power spectrum to find 
conclusive evidence for a non-zero cosmological constant, independent of 
the Hubble diagram of distant Type 1a supernovae. 
Percival et~al. (2002) again used pre-WMAP CMB data and the 
early 2dFGRS power spectrum measurement to place constraints on cosmological 
parameters in flat models. The WMAP team also used the Percival et~al. 
galaxy power spectrum in their estimation of cosmological parameters 
(Spergel et~al. 2003). Other papers have also analyzed the information 
encoded in the 2dFGRS and SDSS power spectra (Tegmark, Zaldarriaga 
\& Hamilton 2001; Pope et~al. 2004; Tegmark et~al. 2004b; 
Seljak et~al. 2005).
In view of the impact of this work, the recent completion by
Cole et~al. (2005) of the power-spectrum analysis of the final
2dFGRS dataset is an important development. The Cole et~al.
results are nearly twice as accurate as those obtained from
the part-complete 2dFGRS in 2001, and a key aim of the current
paper is to see how this affects the outcome of joint analyses
including CMB data. 

In view of these rapid improvements in our knowledge of the cosmological 
parameters, it is also important to take stock of precisely which parts of 
the model are actually being tested. Quite often, restrictive assumptions 
have been adopted for the background cosmology when claims are made about the 
constraints on a particular parameter. 
It is important to establish how robust the constraints really are when 
the data are compared with more general cosmological models. 

Our goal is thus to establish firmly how well the latest CMB and large-scale 
structure (LSS) data 
determine a broad set of cosmological parameters, paying attention to 
how the choice of priors for parameter values and the combination of 
different parameters can influence the results. The outline of the paper is 
as follows. In Section 2, we describe the data used in our 
parameter estimation and set out the various parameter spaces studied. 
In Section 3, we present our main results for the parameter constraints 
obtained by comparing theoretical models to the CMB data and the 
galaxy power spectrum of the final 2dFGRS measured by Cole et~al. (2005). 
We assess the impact of different choices for priors and parameter sets in 
Section 4. 
We explore the justification for using models with different numbers 
of free parameters in Section 5. 
In Section 6, we examine how the parameter constraints change when the 
SDSS galaxy power spectrum measured by Tegmark et~al. (2004b) is used 
instead of the 2dFGRS power spectrum. 
Finally, we summarize our conclusions in Section 7. 

\newpage

\section{The method}

We now set out the approach we will take to constrain the 
values of the basic cosmological parameters. 
In Section~\ref{ssec:data}, we list the CMB and LSS datasets that we 
compare against the theoretical models and explain how these datasets 
are modelled.  The parameter sets that we will consider are defined in 
Section~\ref{ssec:param}. The methodology for searching parameter 
space and placing constraints on parameters is set out in 
Section~\ref{ssec:mechanics}.

\subsection{The datasets} 
\label{ssec:data}

In order to constrain the parameters in our cosmological model, 
we use a compilation of recent measurements of the CMB and 
the power spectrum of galaxy clustering in the local Universe:

\begin{itemize}
\item[(i)] The WMAP first year temperature power spectrum for 
spherical harmonics $2 \le \ell  \le 900$ (Hinshaw et~al. 2003).
\item[(ii)] Observations of the temperature spectrum over the 
spherical harmonic range 
$900 < \ell < 1800$ made up to July 2002 using the Arcminute Cosmology 
Bolometer Array Receiver (ACBAR; Kuo et~al. 2004).
\item[(iii)] The temperature spectrum for $600 < \ell < 1500$ measured using the 
Very Small Array (VSA; Dickinson et~al. 2004).
\item[(iv)] Two years of temperature correlation data with $600 < \ell < 1600$ 
from the Cosmic Background Imager (CBI; Readhead et~al. 2004).
\item[(v)] The WMAP first year temperature-polarization power 
spectrum for spherical harmonics $2 \le \ell \le 450$ (Kogut et~al. 2003).
\item[(vi)] The power spectrum of galaxy clustering measured from the 
final 2dFGRS catalogue (Cole et~al. 2005). 
\end{itemize}

The four measurements (i-iv) of the power spectrum of temperature 
fluctuations in the CMB extend over the spherical 
harmonic range $2 < \ell < 1800$. Some of the available datasets 
extend to higher multipoles. However, we do not include these scales 
in our analysis, as the temperature fluctuations 
on such scales can be strongly affected by secondary sources.  
The WMAP team adopted a similar approach, augmenting 
the first year WMAP data with other experiments which have 
better angular resolution (Spergel et~al. 2003). 
However, the VSA data were not available to the WMAP team 
at the time that the paper by Spergel et~al. was written. 
Theoretical temperature-temperature and temperature-polarization 
spectra are computed for each model using {\tt CAMB} 
(Lewis, Challinor \& Lasenby 2000).  

Cole et~al. (2005) measured the power spectrum of galaxy clustering 
from the final 2dFGRS catalogue. 
The power spectrum measured for galaxies differs in a number of 
ways from the power spectrum for the mass predicted in linear 
perturbation theory: 
(i) Nonlinear evolution of density perturbations leads to coupling 
between Fourier modes, changing the shape of the power spectrum.  
(ii) The galaxy power spectrum is distorted by the gravitationally 
induced peculiar motions of galaxies when a redshift is used to infer 
the distance to each galaxy.  
(iii) The power spectrum of the galaxies could be a modified version 
of the power spectrum of the mass. This phenomenon is known as galaxy bias. 
The ratio between the galaxy and matter spectra could also change with 
scale. (However, we assume a constant bias over the scales considered 
in this paper). 
(iv) The power spectrum measured by Cole et~al. from the 2dFGRS is the
direct transform of the data, and is thus what CMB researchers would
term a pseudo-spectrum. As such, it yields a
convolution of the underlying galaxy power spectrum with the modulus
squared of the Fourier transform of the window
function of the survey.

In order to constrain cosmological parameters, these effects need to 
be modelled. The accuracy of the modelling requires that 
the comparison between theory and observation should be restricted 
to a limited range of scales. 
We use the 2dFGRS power spectrum data for $k<0.15 \,h \mathrm{{Mpc^{-1}}}$ 
and discard measurements with $k< 0.02 \,h \mathrm{{Mpc^{-1}}}$
which could be affected by uncertainties in the mean density of 
galaxies within the survey. 
We follow the scheme used by Cole et~al. who applied a correction 
for non-linearity and scale-dependent bias to the shape of $P(k)$ of the form 
\begin{equation}
P_{\rm gal}(k)=b^{2}\frac{1+Qk^{2}}{1+Ak}\,P_{\rm lin}(k),  \label{eq:q+a}
\end{equation}
where $A=1.4$ and $Q=4.6$ are the preferred values and b is a constant bias 
factor. 
This formula is deduced by comparison with detailed numerical
galaxy-formation models: these show that the value of $A$ is robust, but
the exact value of $Q$ depends on galaxy type and also has some
uncertainty depending on how the modelling is done. These results were
used to determine a range of allowed values for $Q$, from which the value
$Q=4.6$ is preferred; with this choice, robust parameter constraints are
obtained if one considers maximum $k$ values beyond our limit of
$0.15\,h\,{\rm Mpc}^{-1}$. For this limit, neglecting the correction
entirely and simply fitting linear theory yields almost identical results
to those presented here. In particular, it has no impact on the marginal
indication of a deviation from $n_{\rm s}=1$.

\subsection{The parameter space}
\label{ssec:param}

In this paper, we make the basic assumption that the primordial 
density fluctuations were adiabatic, Gaussian and had a power law 
spectrum of Fourier amplitudes. As pointed out by Leach \& Liddle 
(2003a), the CMB data prior to WMAP were of insufficient quality 
to justify the rejection of this simple hypothesis. Following 
the release of the WMAP first year results, which do have the 
precision required to test this model, our assumptions remain 
well motivated. Komatsu et al. (2003) found that the WMAP sky maps 
are consistent with Gaussian primordial fluctuations to a much higher 
precision than was attainable with COBE. Peiris et al. (2003) 
found that models with a spectral index varying slowly with wavenumber 
give slightly better fits to the WMAP data, particularly when combined 
with estimates of the power spectrum of the Lyman-$\alpha$ forest. 
However, the evidence for a running spectral index is weak and has been 
disputed by other groups (e.g. Bridle et~al. 2003b; Slosar, Seljak \& 
Makarov 2003; Seljak et~al. 2005). 
Bennett et~al. (2003) and Spergel et~al. (2003) point out that,  
on large scales, a few modes of the CMB temperature power spectrum 
measured by WMAP lie below the predictions of the standard $\Lambda$CDM 
model. 
One interpretation of this apparent discrepancy is that new physics 
may be needed (e.g. Bridle et~al. 2003a; Efstathiou 2003). However, several 
studies have argued that the disagreement is actually less significant 
than was first claimed (Gazta\~{n}aga et~al. 2003; 
de Oliveira-Costa et~al. 2003; Efstathiou 2004). 

>From the above starting point, the cosmological model we consider 
is defined by eleven parameters: 
\begin{equation}
\mathbf{P\equiv (}\Omega_{k},\omega_{\rm dm},\omega
_{\rm b},f_{\rm \nu},w_{\rm DE},\tau ,n_{\rm s},A_{\rm s},r,b,\Theta).  
\label{eq:param}
\end{equation}
There are eight further basic quantities whose values can be derived 
given the above set:  
\begin{equation}
\mathbf{P_{\rm derived} \equiv (}
\Omega_{\rm DE},h,\Omega_{\rm m},\sigma_{8},z_{\rm re},t_{0},\sum m_{\nu}, n_{\rm t}
).  
\label{eq:paramderived}
\end{equation}
We now go through the parameters in these lists, defining each one and 
explaining how the values of the derived parameters are obtained. 

The are five quantities that describe the homogeneous background
cosmology through various contributions to the mass-energy density. 
These are, in units of the critical density: $\Omega_{k}$, which describes 
the curvature of the universe; $\Omega_{\rm DE}$, the energy-density of the 
dark energy; $\omega_{\rm dm}\equiv \Omega_{\rm dm}h^{2}$, the density of 
the dark matter (where $\Omega_{\rm dm}=\Omega_{\rm cdm}+ 
\Omega_{\nu }$ is the sum of the cold and hot dark matter 
components and $h$ is Hubble's constant in units of 
$100 \,{\rm km s}^{-1}{\rm Mpc}^{-1}$); $\omega_{\rm b}\equiv 
\Omega_{\rm b}h^{2}$,  the baryon density; and 
$f_{\rm \nu }=\Omega_{\rm \nu}/\Omega_{\rm dm },$ the fraction of the 
dark matter in the form of massive neutrinos. 
The sum of neutrino masses is given by $\sum m_{\nu}$.
The matter density parameter is given by $\Omega_{\rm m} = \Omega_{\rm dm} + 
\Omega_{\rm b}$. 
The value of the Hubble constant is derived from $h = \sqrt{(\omega_{\rm dm} 
+ \omega_{\rm b})/\Omega_{\rm m}}$.
The energy density of the dark energy is set by $\Omega_{\rm DE} = 
1 - \Omega_{\rm m} - \Omega_{k}$. 
The dark energy component is assumed to have an equation of state that 
is independent of redshift, with the ratio of pressure to density given 
by $w_{\rm DE}$. 

There are four quantities that describe the form of the initial 
fluctuations; the spectral indices, $n_{\rm s}$ and $n_{\rm t}$, and the 
primordial amplitudes, $A_{\rm s}$ and $rA_{\rm s}$, of scalar 
and tensor fluctuations respectively. These parameter values are quoted at 
the `pivot' scale wavenumber of $k=0.05\, {\rm Mpc}^{-1}$.  
We can translate the results obtained for $A_{\rm s}$ into a constraint 
on the more familiar parameter $\sigma_{8}$, the {\it rms} linear 
perturbation theory variance in spheres of radius $8~h^{-1}$Mpc,
using the matter fluctuation transfer function. 
Note that when we consider tensor modes, we make the slow-roll assumption 
that $n_{\rm t}=-r/8$. 

The bias factor, $b\equiv \sqrt{P_{\rm gal}(k)/P_{\rm DM}(k)}$, 
describes the difference in amplitude between the galaxy power 
spectrum and that of the underlying dark matter. 
The value of $b$ is marginalized over, using the analytic expression 
given in Appendix F of Lewis \& Bridle (2002). 
We assume that the reionization of the neutral intergalactic medium 
occurred instantaneously, with an optical depth given by $\tau$; the 
redshift of reionization, $z_{\rm re}$, depends upon a combination 
of parameters (see Table 1 of Tegmark et~al. 2004b). 
The age of the universe is $t_{0}$.

Finally, $\Theta$ gives the ratio of the sound horizon scale 
at the epoch of decoupling to the angular diameter distance 
to the corresponding redshift and replaces the Hubble constant  
as a base parameter (Kosowsky, Milosavljevic \& Jimenez 2002). 
We have chosen to use this parameter, rather than, for example, the 
energy density in dark energy, since it has a posterior 
distribution that is close to Gaussian. 
This reduces degeneracies between parameters and results in a faster 
convergence of our search of parameter space (see Section~\ref{ssec:mechanics}), 
compared with studies in which parameters such as $\Omega_{\rm DE}$, 
which does not have a Gaussian posterior distribution,  are allowed to vary.  
This approach is a standard practice even though, usually, the 
final results are expressed in terms of more familiar parameters 
such as $\Omega_{\rm DE}$ or $h$.
However, care must be taken when comparing our results 
with those from studies which have assumed flat priors on different 
parameters in their Bayesian analysis. Such choices may affect 
the final results in a way that is difficult to quantify.

We do not attempt to vary all eleven parameters of the model at once. 
Such an approach would lead to a mixture of poor estimates of 
the values of individual parameters and constraints on various 
combinations of parameters. 
As we are primarily interested in deriving the best possible 
constraints on individual parameters, we instead consider 
subsets of parameter space, varying five, six or seven 
parameters at a time. Of the remaining parameters, some are 
held at fixed values and the others are referred 
to as derived parameters. The values of the derived parameters follow 
from the values of other parameters, once the assumptions 
made in each parameter space have been taken into account. 
We will now set out each of our parameter spaces in turn, stating 
which parameters are varied and which are held fixed. In all cases, 
the bias parameter, $b$, is marginalized over, so we do not include 
this in the list of parameters whose values are constrained.  

In the simplest case, we vary five parameters which we refer to as 
the `basic-five' (b5) parameter set. The following parameters are 
allowed to float: 
\begin{equation}
\mathbf{
P^{5}_{\rm varied} \equiv (}
\omega_{\rm dm},
\omega_{\rm b},
\tau,
A_{\rm s},
\Theta
).  \label{eq:param5}
\end{equation}
The values of the fixed parameters in the basic-five model are: 
\begin{equation}
\mathbf{P^{5}_{\rm fixed} \equiv (}\Omega_{\rm k}=0,f_{\rm \nu}=0,
w_{\rm DE}=-1,n_{\rm s}=1,r=0).  \label{eq:paramf5}
\end{equation}
The results of this model are discussed in Section 3.2. 

The basic-five set is expanded to allow the value of the 
scalar spectral index to float, giving the basic-six (b6) model (see Section~3.3): 
\begin{equation}
\mathbf{
P^{6}_{\rm varied} \equiv (}
\omega_{\rm dm},
\omega_{\rm b},
\tau,
n_{\rm s},
A_{\rm s},
\Theta
).  \label{eq:param6}
\end{equation}
The fixed parameters in the basic-six model are: 
\begin{equation}
\mathbf{P^{6}_{\rm fixed} \equiv (}\Omega_{\rm k}=0,f_{\rm \nu}=0,
w_{\rm DE}=-1,r=0).  \label{eq:paramf6}
\end{equation}

\begin{table}
\begin{center}
\caption{ The parameter space probed in our analysis. We assume a flat 
prior in each case. We do not vary the values of all parameters at the 
same time; the parameter spaces that we consider are set out 
in Section~\ref{ssec:param}. 
}
\end{center}
\begin{center}
\begin{tabular}[t]{cc}
\hline\hline
Parameter       & Allowed range \\ \hline\hline
$\Omega_{k}$   & $-$0.3 -- 0.3   \\ 
$\omega_{\rm dm}$  & 0.01 -- 0.99   \\ 
$\omega_{\rm b}$   & 0.005 -- 0.1  \\ 
$f_{\rm \nu }$      & 0 -- 0.5   \\ 
$w_{\rm DE }$       & $-$2. -- 0  \\ 
$\tau $             & 0 -- 0.8   \\ 
$n_{\rm s}$         & 0.5 -- 1.5    \\ 
$\log_{10}(10^{10}A_{\rm s})$ & 2.7 -- 4.0 \\ 
$r$             & 0 -- 1  \\ 
$\Theta  $      & 0.5 -- 10   \\ 
\hline\hline
\end{tabular}
\end{center}
\label{tab:param}
\end{table}

We also consider four parameter spaces in which one additional parameter is 
constrained along with the basic-six set. 
In Section~3.3, the additional parameter is the mass fraction of massive neutrinos, 
$f_{\nu}$: 
\begin{equation}
\mathbf{
P^{6+f_{\nu}}_{\rm varied} \equiv (}
\omega_{\rm dm},
\omega_{\rm b},
f_{\nu}, 
\tau,
n_{\rm s},
A_{\rm s},
\Theta
).  \label{eq:param7a}
\end{equation}
The fixed parameters in this case are: 
\begin{equation}
\mathbf{P^{6 + f_{\nu}}_{\rm fixed} \equiv (}\Omega_{\rm k}=0,
w_{\rm DE}=-1,r=0).  \label{eq:paramf7a}
\end{equation}
In Section~3.4, the curvature of the universe, $\Omega_{k}$ is allowed to float and 
the fraction of massive neutrinos is once again held fixed: 
\begin{equation}
\mathbf{
P^{6+\Omega_{k}}_{\rm varied} \equiv (}
\Omega_{ k},
\omega_{\rm dm},
\omega_{\rm b},
\tau,
n_{\rm s},
A_{\rm s},
\Theta
).  \label{eq:param7b}
\end{equation}
\begin{equation}
\mathbf{P^{6 + \Omega_{k}}_{\rm fixed} \equiv (}
f_{\nu}=0,w_{\rm DE}=-1,r=0).  \label{eq:paramf7b}
\end{equation}
In Section~3.5, the equation of state of the dark energy is varied: 
\begin{equation}
\mathbf{
P^{6+w_{\rm DE}}_{\rm varied} \equiv (}
\omega_{\rm dm},
\omega_{\rm b},
w_{\rm DE},
\tau,
n_{\rm s},
A_{\rm s},
\Theta
).  \label{eq:param7c}
\end{equation}
In this case, the fixed parameters are: 
\begin{equation}
\mathbf{P^{6 + w_{\rm DE}}_{\rm fixed} \equiv (}\Omega_{\rm k}=0,
f_{\nu}=0,r=0).  \label{eq:paramf7c}
\end{equation}
Finally, in Section~3.6, the constraints on tensor modes are investigated: 
\begin{equation}
\mathbf{
P^{6+r}_{\rm varied} \equiv (}
\omega_{\rm dm},
\omega_{\rm b},
\tau,
n_{\rm s},
A_{\rm s},
r,
\Theta
),  \label{eq:param7d}
\end{equation}
with the fixed parameters given by: 
\begin{equation}
\mathbf{P^{6 + r}_{\rm fixed} \equiv (}\Omega_{\rm k}=0,
f_{\nu}=0,w_{\rm DE}=-1).  \label{eq:paramf7d}
\end{equation}
Table~1 summarizes the ranges considered for 
different cosmological parameters when their values are allowed 
to vary.

\subsection{Constraining parameters} 
\label{ssec:mechanics}

The prohibitive computational cost of generating CMB power spectra and 
matter transfer functions for all the grid points in a multidimensional 
parameter space has driven the development of codes that sample the space 
selectively, guided by the shape of the likelihood surface. We use a 
Markov Chain Monte Carlo (MCMC) approach to search the parameter space of 
the cosmological model (for a recent example of the application of the MCMC 
algorithm to cosmological applications, see Percival 2004). In brief, this 
algorithm involves conducting a series of searches of parameter space called 
chains. The chains are started at widely separated locations within the 
space. The next link in a chain is made in a randomly chosen direction in 
the parameter space. The new link becomes part of the chain if it passes a 
test devised by Metropolis et~al. (1953); in summary, links for which the 
likelihood increases are always retained, otherwise acceptance 
occurs with a probability that is
the ratio of likelihoods between the new and old links. 
If a link is rejected, a new randomly generated step is taken in the 
parameter space. 
This rate of hopping between pairs of points in parameter space
satisfies the principle of detailed balance, so that the chains
should asymptotically take up a stationary probability 
distribution that follows the likelihood surface.
The advantage of this method is that marginalization 
(i.e. integration of the posterior distribution over uninteresting parameters) 
is extremely easy: one simply adds up
the number of links that fall within binned intervals of 
the interesting parameter values (see the appendices in Lewis \& Bridle 2002). 

The results presented in this paper were generated with the 
publicly available CosmoMC code of Lewis \& Bridle (2002). 
We have compared the parameter constraints obtained with 
this code with those from an 
independent code written by one of us (WJP), and find excellent agreement 
between the two sets of results. CosmoMC uses the {\tt CAMB} package to compute 
power spectra for the CMB and matter fluctuations (Lewis, Challinor \& 
Lasenby 2000). Our analysis was carried out in parallel on the Cosmology 
Machine at Durham University. For each parameter set considered, we ran 
twenty separate chains using the Message Passing Interface (MPI) convergence 
criterion to stop 
the chains when the Gelman and Rubin (1992) statistic $R<1.02$, 
which is a significantly more stringent criterion than is usually adopted 
(Verde et~al. 2003; Seljak et~al. 2005). 
The length of chain generated before the above convergence criterion 
is achieved depends upon the datasets used.  For CMB data alone, the 
chains typically 
have on the order of 10,000 links; in the case of CMB plus the 2dFGRS $P(k)$, 
convergence can be reached more quickly.  
In total, our calculations have accounted for the equivalent of more 
than 30 CPU years on a single processor.

\section{Results}
\begin{figure*}
\centering
\centerline{\includegraphics[width=0.9\textwidth]{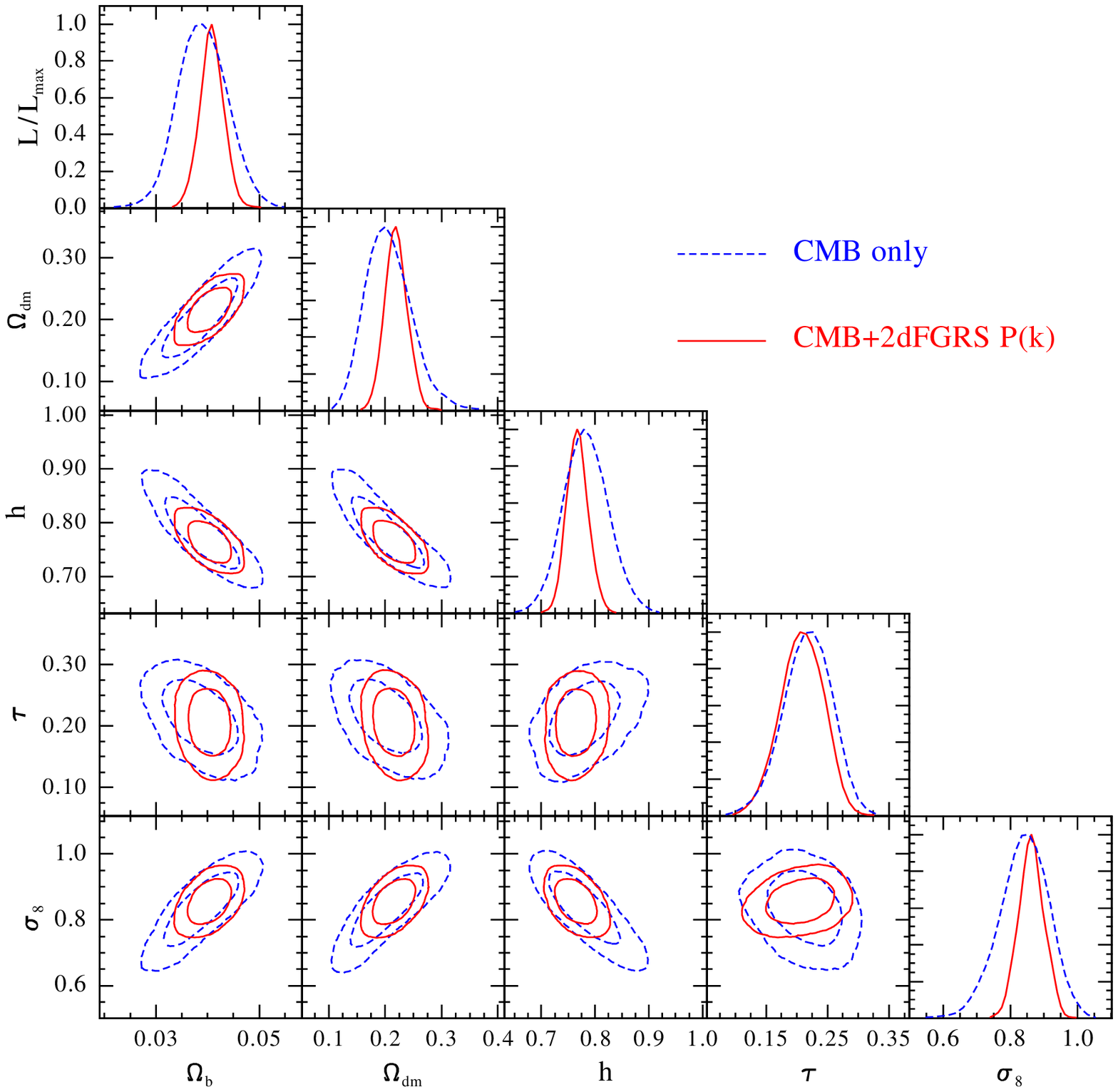}}
\caption{
Marginalized posterior likelihoods for the cosmological parameters 
in the basic-five model determined from CMB information only (dashed lines) 
and CMB+2dFGRS $P(k)$ (solid lines). 
The diagonal shows the likelihood for individual parameters; the other 
panels show the likelihood contours for pairs of parameters, marginalizing 
over the other parameters. The contours show 
$-2\Delta \ln(L/L_{\rm max}) = 2.3$ and $6.17$.
}
\label{fig:b5-san}
\end{figure*}

\begin{figure*}
\centering
\centerline{\includegraphics[width=0.9\textwidth]{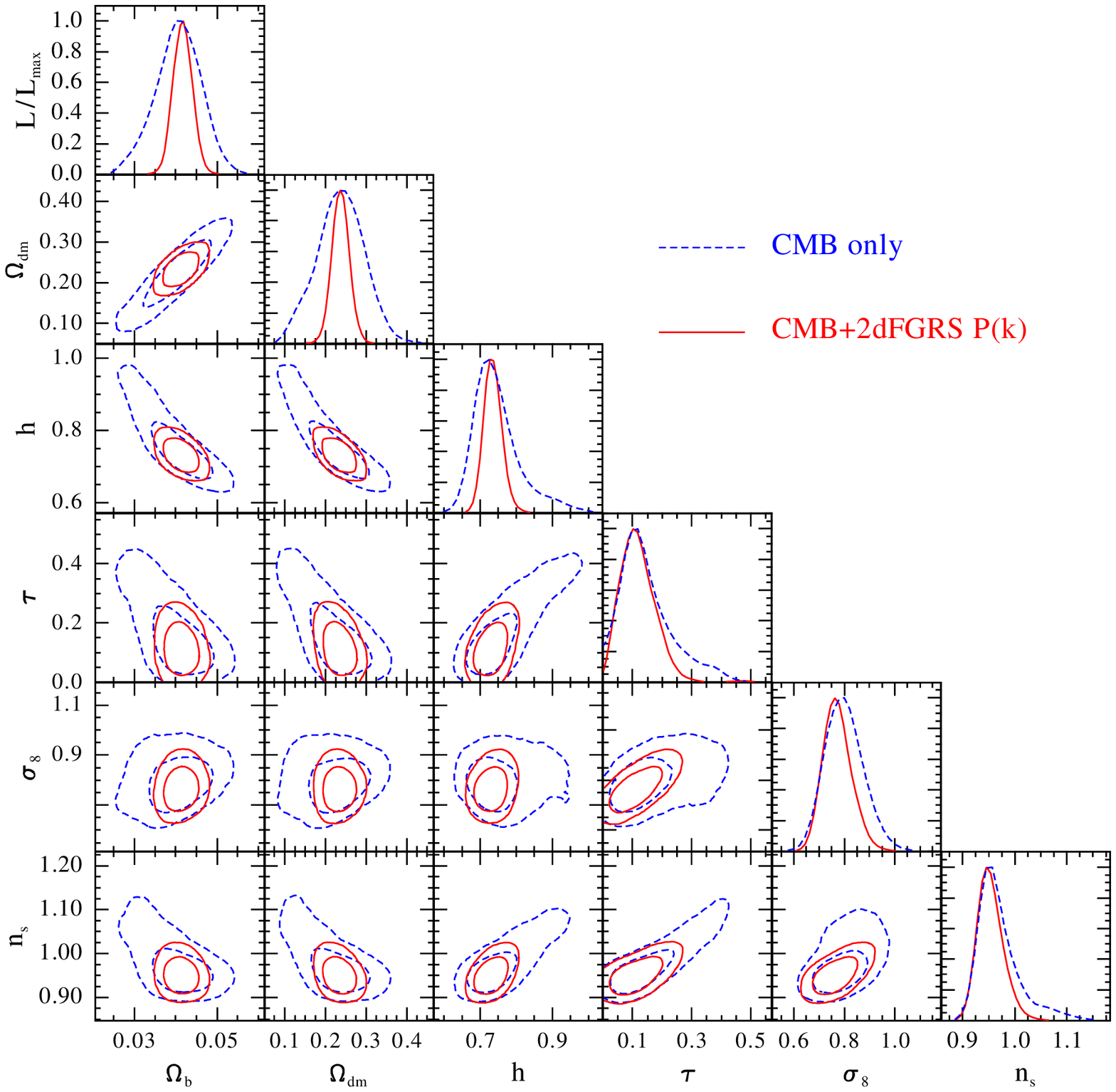}}
\caption{
Marginalized posterior likelihoods for the cosmological parameters 
in the basic-six model determined from CMB information only (dashed lines) 
and CMB plus 2dFGRS $P(k)$ (solid lines).
}
\label{fig:b6-san}
\end{figure*}

In this Section, we carry out a systematic study of the constraints 
placed on the values of cosmological parameters by the CMB and LSS 
datasets listed in Section~\ref{ssec:data}. We vary three aspects of the comparison: 
(i) The datasets used. We compare constraints obtained from the 
CMB data alone (Table 2) with those obtained from the CMB data in combination 
with the 2dFGRS power spectrum (Table 3). This allows us to see which 
parameters are constrained more strongly when the CMB data are combined  
with a measurement of the galaxy power spectrum. 
(ii) The number of parameters varied. We consider models in which 5, 6 
or 7 parameters are allowed to float whilst the other parameters are 
held at fixed values (see Section~\ref{ssec:param} for the definition of 
our parameter spaces). 
(iii) The combination of parameters. In our 7 parameter models, 
we add one additional parameter to our basic-six set (Eq.~\ref{eq:param6}) 
and explore how different choices for this additional parameter can affect 
the parameter constraints. 

Our results are summarized in Tables 2 and 3. 
In the top half of each table, we show the values of the fundamental 
parameters. These are either the range of values derived by comparison 
with a particular dataset or the value that a parameter is fixed at in the 
analysis, as explained in Section~\ref{ssec:param}. 
In the lower part of the tables, we quote the values of other useful 
parameters (as listed in Eq.~\ref{eq:paramderived}). 
These parameters are not varied directly in our analysis. However, 
their values can be derived from the results in the upper half of the 
table, as explained in Section~2.2.

In Section 3.1, we present the results for a minimal cosmological model 
with five parameters, the basic-five set. 
In Section 3.2, we consider six parameters, the basic-six set, 
allowing the spectral index of scalar fluctuations to float. 
Sections 3.3 to 3.6 are devoted to seven parameter models, with different 
choices for the `final' parameter that augments the basic-six set, as follows: 
Section~3.3 the mass fraction of massive neutrinos, $f_{\rm \nu}$, 
Section~3.4 non-flat models, 
Section~3.5 the dark energy equation of state, $w_{\rm DE}$, 
Section~3.6 the addition of tensor perturbations. 

In the results tables, unless otherwise stated, we quote errors that 
enclose $68\%$ of the probability around the mean value of each parameter.
In the subsequent figures showing the marginalized posterior likelihood 
surface for two parameters, the contours mark the locus 
where $ -2 \Delta \ln \left({\cal L} / {\cal L}_{\rm max}\right) 
= 2.30$ and $6.17$, corresponding to the 68\% and 95\% limits 
respectively; for the case of a Gaussian likelihood, these contours 
correspond to the `1$\sigma$' and `2$\sigma$' limits for two 
degrees of freedom.

\begin{table*}
\renewcommand\arraystretch{1.45}
\begin{center}
\caption{Marginalized 68\% interval constraints (unless stated otherwise) on cosmological parameters 
obtained using CMB information only for the different hypothesis and 
parameter sets analysed. 
The models are defined in Section~2.2.}
\begin{tabular}{lcccccc}
\hline\hline
& b5 & b6 & b6 + $f_{\nu}$  & b6 + $\Omega_{k}$ & b6 + $w_{\rm DE}$ & b6 + $r$ \\ 
\hline\hline
$\Omega_{k}$        & 0                            & 0                            & 0                            &  
                       $-0.074^{+0.049}_{-0.052}$   & 0                            & 0                       \\ 
$\Theta$             & $1.0449^{+0.0041}_{-0.0042}$ & $1.0420^{+0.0052}_{-0.0052}$ & $1.0428^{+0.0059}_{-0.0058}$ &
                       $1.0427^{+0.0063}_{-0.0062}$ & $1.0426^{+0.0052}_{-0.0052}$ & $1.0433^{+0.0051}_{-0.0051}$  \\ 
$\omega_{\rm dm}$        & $0.101^{+0.011}_{-0.011}$   & $0.105^{+0.013}_{-0.013}$    & $0.113^{+0.014}_{-0.015}$    &
                       $0.095^{+0.019}_{-0.026}$    & $0.105^{+0.013}_{-0.013}$    & $0.099^{+0.010}_{-0.011}$  \\ 
$\omega_{\rm b}$        & $0.0239^{+0.0007}_{-0.0007}$ & $0.0229^{+0.0012}_{-0.0013}$ & $0.0226^{+0.0015}_{-0.0016}$ &
                       $0.0238^{+0.0032}_{-0.0022}$ & $0.0231^{+0.0013}_{-0.0013}$ & $0.0236^{+0.0013}_{-0.0013}$  \\ 
$f_{\nu }$           & 0                            & 0                            & $<0.182$ (95\%)    &
                       0                            & 0                            & 0 \\ 
$\tau $              & $0.217^{+0.037}_{-0.036}$    & $0.150^{+0.084}_{-0.078}$    & $0.161^{+0.101}_{-0.091}$    &
                       $0.24^{+0.24}_{- 0.16}$      & $0.142^{+0.074}_{-0.073}$    & $0.126^{+0.062}_{-0.062}$ \\ 
$w_{\rm DE}$             & $-1$                          & $-1$                          & $-1$                          &
                      $-1$                          & $-0.93^{+0.49}_{-0.47}$      & $-1$ \\ 
$n_{\rm s}$              & 1                            & $0.970^{+0.033}_{-0.033}$    & $0.957^{+0.045}_{-0.047}$    &
                       $1.00^{+0.11}_{-0.07}$       & $0.974^{+0.038}_{-0.037}$    & $0.994^{+0.033}_{-0.033}$ \\ 
$\log_{10}(10^{10}A_{\rm s})$ & $3.270^{+0.059}_{-0.058}$    & $3.14^{+0.16}_{-0.15}$       & $3.14^{+0.19}_{-0.18}$       &
                       $3.29^{+0.42}_{-0.28}$       & $3.12^{+0.14}_{-0.14}$       & $3.07^{+0.13}_{- 0.13}$  \\ 
$r$                  & 0                            & 0                            & 0                            &
                       0                            & 0                            & $<0.52$ (95\%)\\\hline
$\Omega_{\rm DE}$       & $0.793^{+0.039}_{-0.038}$    & $0.762^{+0.056}_{-0.055}$    & $0.68^{+0.10}_{-0.10}$       &
                       $0.63^{+0.18}_{-0.17}$       & $0.71^{+0.12}_{-0.14}$       & $0.798^{+0.041}_{-0.042}$  \\
$t_{0}/{\rm Gyr}$          & $13.38^{+0.12}_{-0.12}$      & $13.58^{+0.26}_{-0.25}$      & $14.03^{+0.47}_{-0.44}$      &
                       $16.3^{+1.4}_{-1.5}$         & $13.79^{+0.50}_{-0.45}$      & $13.43^{+0.25}_{-0.26}$  \\ 
$\Omega_{\rm m}$           & $0.207^{+0.038}_{-0.039}$    & $0.237^{+0.055}_{-0.056}$    & $0.32^{+0.10}_{-0.10}$       &
                       $0.44^{+0.21}_{-0.20}$       & $0.28^{+0.14}_{-0.12}$       & $0.202^{+ 0.041}_{-0.042}$  \\
$\sigma_{8}$           & $0.840^{+0.069}_{-0.069}$    & $0.800^{+0.073}_{-0.072}$    & $0.63^{+0.12}_{-0.12}$    &
                       $0.776^{+0.076}_{-0.072}$    & $0.75^{+0.18}_{-0.18}$       & $0.706^{+0.093}_{-0.097}$  \\
$z_{\rm re}$             & $19.6^{+2.1}_{-2.1}$         & $15.0^{+5.6}_{-5.1}$         & $15.9^{+6.8}_{-5.9}$         &
                       $18.6^{+9.7}_{-7.7}$         & $14.5^{+5.1}_{-5.0}$         & $15.4^{+5.3}_{-5.3}$  \\
$h$                  & $0.783^{+0.040}_{-0.040}$    & $0.747^{+0.055}_{-0.056}$    & $0.674^{+0.078}_{-0.082}$    &
                       $0.54^{+0.11}_{-0.11}$       & $0.72^{+0.18}_{-0.17}$       & $0.786^{+0.053}_{-0.052}$  \\
$\sum{m_{\nu}}/\rm{eV}$ & 0                         & 0                            & $<2.09$                     &
                       0                            & 0                            & 0                          \\
\hline\hline
\end{tabular}
\end{center}
\end{table*}

\begin{table*}
\renewcommand\arraystretch{1.45}
\begin{center}
\caption{Marginalized 68\% interval constraints (unless stated otherwise) on cosmological parameters 
obtained using information from CMB and the 2dFGRS power spectrum for 
the different hypothesis and parameter sets analysed. 
The models are defined 
in Section~2.2.}
\begin{tabular}{lcccccc}
\hline\hline
& b5 & b6 & b6 + $f_{\nu}$  & b6 + $\Omega_{k}$ & b6 + $w_{\rm DE}$ & b6 + $r$ \\ 
\hline\hline
$\Omega_{k}$        & 0                            & 0                            & 0                            &
                       $-0.029^{+0.018}_{-0.018}$   & 0                            & 0                           \\ 
$\Theta$             & $1.0453^{+0.0038}_{-0.0037}$ & $1.0403^{+0.0046}_{-0.0045}$ & $1.0411^{+0.0050}_{-0.0046}$ &
                       $1.0458^{+0.0079}_{-0.0076}$ & $1.0422^{+0.0055}_{-0.0055}$ & $1.0425^{+0.0049}_{-0.0049}$\\ 
$\omega_{\rm dm}$       & $0.1046^{+0.0055}_{-0.0053}$ & $0.1051^{+0.0046}_{-0.0047}$ & $0.1100^{+0.0062}_{-0.0067}$ &
                       $0.083^{+0.015}_{-0.015}$    & $0.097^{+0.011}_{-0.011}$    & $0.1037^{+0.0050}_{-0.0050}$  \\ 
$\omega_{\rm b}$        & $0.0240^{+0.0006}_{-0.0006}$ & $0.0225^{+0.0010}_{-0.0010}$ & $0.0224^{+0.0012}_{-0.0011}$ &
                       $0.0252^{+0.0033}_{-0.0030}$ & $0.0233^{+0.0016}_{-0.0016}$ & $0.0233^{+0.0011}_{-0.0011}$ \\ 
$f_{\nu }$           & 0                            & 0                            & $<0.105$ (95\%)   &
                       0                            & 0                            & 0                           \\ 
$\tau $              & $0.208^{+0.034}_{- 0.034}$   & $0.118^{+0.057}_{-0.056}$    & $0.143^{+0.076}_{-0.071}$    &
                       $0.33^{+0.18}_{-0.19}$       & $0.174^{+0.107}_{-0.095}$    & $0.109^{+0.053}_{-0.053}$   \\ 
$w_{\rm DE}$             & $-1$                          & $-1$                          & $-1$                          &
                      $-1$                          & $-0.85^{+0.18}_{-0.17}$      & $-1$                         \\ 
$n_{\rm s}$              & 1                            & $0.954^{+0.023}_{-0.023}$    & $0.957^{+0.031}_{-0.029}$    &
                       $1.05^{+0.10}_{- 0.10}$      & $0.985^{+0.053}_{-0.046}$    & $0.979^{+0.028}_{-0.028} $   \\ 
$\log_{10}(10^{10}A_{\rm s})$ & $3.268^{+0.060}_{-0.060}$    & $3.06^{+0.12}_{-0.12}$       & $3.11^{+0.15}_{-0.14}$       &
                       $3.44^{+0.35}_{-0.37}$       & $3.16^{+0.20}_{-0.18}$       & $3.05^{+0.11}_{- 0.11}$     \\ 
$r$                  & 0                            & 0                            & 0                            &
                       0                            & 0                            & $<0.41$  (95\%)    \\ \hline
$\Omega_{\rm DE}$       & $0.781^{+0.019}_{-0.020}$    & $0.763^{+0.020}_{-0.020}$    & $0.718^{+0.042}_{-0.037}$    &
                       $0.796^{+0.040}_{-0.040}$    & $0.759^{+0.024}_{-0.024}$    & $0.778^{+0.021}_{-0.022}$   \\
$t_{0}/{\rm Gyr}$          & $13.39^{+0.11}_{-0.11}$      & $13.69^{+0.19}_{-0.20}$      & $13.94^{+0.26}_{-0.26}$      &
                       $14.97^{+0.77}_{-0.79}$      & $13.70^{+0.26}_{-0.26}$      & $13.54^{+0.23}_{-0.23}$     \\
$\Omega_{\rm m}$           & $0.219^{+0.020}_{-0.019}$    & $0.237^{+0.020}_{-0.020}$    & $ 0.282^{+0.037}_{-0.042}$    &
                       $0.234^{+0.028}_{-0.027}$    & $0.241^{+0.024}_{-0.024}$    & $0.224^{+0.022}_{-0.022}$   \\
$\sigma_{8}$           & $0.863^{+0.037}_{-0.037}$    & $0.773^{+0.054}_{-0.053}$    & $0.678^{+0.073}_{-0.072}$    &
                       $0.817^{+0.077}_{-0.079}$    & $0.711^{+0.098}_{-0.099}$    & $0.769^{+0.053}_{-0.062}$  \\
$z_{\rm re}$             & $19.2^{+2.1}_{-2.1}$         & $13.1^{+4.3}_{-4.3}$         & $15.1^{+5.2}_{-5.1}$         &
                       $22.6^{+6.2}_{-7.9}$         & $16.1^{+6.2}_{-5.8}$         & $12.1^{+4.1}_{-4.2}$        \\
$h$                  & $0.776^{+0.020}_{-0.019}$    & $0.735^{+0.022}_{-0.023}$    & $0.691^{+0.038}_{-0.038}$    &
                       $0.684^{+0.035}_{-0.035}$    & $0.708^{+0.062}_{-0.058}$    & $0.755^{+0.028}_{-0.029}$   \\
$\sum{m_{\nu}}/\rm{eV}$ & 0                         & 0                            & $<1.16$ (95\%) &
                       0                            & 0                            & 0                          \\
\hline\hline
\end{tabular}
\end{center}
\end{table*}

\subsection{The simplest case -- five parameters}


We first concentrate on the simplest possible model that gives an accurate 
description of the data sets, the basic-five parameter space defined by 
Eqs.~\ref{eq:param5} and \ref{eq:paramf5}. This model does a 
remarkably good job of reproducing the CMB data, 
with tight constraints  
obtained on the values of the subset of five cosmological 
parameters varied, as shown by the dashed lines in Fig.~\ref{fig:b5-san} 
and column 2 of Table 2. It is clear from Fig.~\ref{fig:b5-san} and Table 3, 
that, when the 2dFGRS $P(k)$ is included, the results show an impressive 
consistency with those obtained from the CMB data alone. For example, 
in the case of the physical density of dark matter, $w_{\rm dm}$, the 
central values derived when comparing to CMB data alone and 
to CMB plus 2dFGRS agree well within the uncertainties. 
However, in a number of cases, there is a significant improvement in the 
parameter constraints obtained when the 2dFGRS $P(k)$ data are included. 
For example, the range of $w_{\rm dm}$ values derived is narrower by a 
factor of 2 when the 2dFGRS $P(k)$ is included in the fit, as the LSS
data breaks the horizon-angle degeneracy arising from 
CMB models with the same position of the first peak in the angular 
power spectrum (e.g. Percival et~al. 2002).
A similar reduction in uncertainty occurs for the derived 
parameters $\sigma_{8}$ and $h$. The CMB power spectrum is 
sensitive to the parameter $\omega_{\rm dm}=\Omega_{\rm dm}h^{2}$ 
whereas the matter $P(k)$ depends on the parameter combination 
$\Omega_{\rm dm}h$. The incorporation of $P(k)$ into the analysis 
helps to break the degeneracy between $\Omega_{\rm dm}$ and $h$ present 
in the theoretical predictions for the CMB, thus tightening the
constraints on these parameters, as well as on $\omega_{\rm dm}$. 
Cole et~al. (2005) used the 2dFGRS $P(k)$ to place constraints on the 
parameter combinations $\Omega_{\rm m}h$ and $\Omega_{\rm b}/\Omega_{\rm m}$, 
and, in conjunction with the WMAP temperature power spectrum, on 
$\Omega_{\rm m}$.  The model that Cole et~al. considered is a restricted 
version of our basic-five model (they assumed $h=0.72$). It is reassuring to 
note that our results are in excellent agreement with those obtained by 
Cole et~al.; in particular, we confirm their finding of a matter density 
significantly below the canonical $\Lambda$CDM value of $\Omega_{\rm m}=0.3$.
The success of this simple model in describing the current CMB and 
LSS data is remarkable. This `minimalist model' does a 
perfectly good job of accounting for the form of the most precise probes 
of the cosmological world model that are available to us today. 

\subsection{Six parameters -- including the scalar spectral index}

We now expand our model to allow variations in the scalar 
spectral index, $n_{\rm s}$, which we call the `basic-six'  
parameter space (defined by Eqs.~\ref{eq:param6} and~\ref{eq:paramf6}). 
Fig.~\ref{fig:b6-san} shows the marginalized 
likelihoods for this parameter set (along the diagonal), together 
with the two dimensional likelihood contours for different combinations 
of parameters. 
The results are shown using the CMB data alone (dashed lines) 
and for CMB plus 2dFGRS $P(k)$ (solid lines). 
The additional degree of freedom gives rise to a well known 
degeneracy that involves all six parameters and which is seen 
most clearly in the optical depth to last scattering, $\tau$, and the spectral 
index and amplitude of scalar fluctuations, $n_{\rm s}$ and $A_{\rm s}$ 
respectively. 
This degeneracy leads to the production of similar power spectra as 
the parameter values, with the exception of $\omega_{\rm dm}$, are increased 
(see Tegmark et al. 2004b for a full description of how the degeneracy 
works in practice).
Table 2 shows that in the case of the CMB data alone, 
the results for the best fitting parameters in the basic-six case 
are, for the most part, very similar to those obtained for the 
basic-five parameter set. The two exceptions are $\tau$ and $A_{\rm s}$, 
for which slightly lower values are obtained in the basic-six case. 
This is also a consequence of the above degeneracy, since, as the data 
prefer $n_{\rm s}<1$, the best fitting values for $\tau$ and $A_{\rm s}$ 
also decrease. Another consequence of the degeneracy is to broaden 
the allowed regions compared with those obtained for the basic-five 
parameter set. The 2dFGRS power spectrum helps to break this degeneracy, 
particularly by tightening the constraints on $w_{\rm dm}$. The results 
listed in column 3 of Table 3 show that the marginalized constraints 
obtained in this case are in complete agreement with those in the CMB only 
case, but with tighter allowed ranges. This reinforces the consistency 
of the results obtained from CMB alone and CMB plus 2dFGRS $P(k)$ that we 
found in the basic-five case.

One particularly remarkable result is the recovered value 
of the spectral index of 
scalar perturbations, $n_{\rm s}$. 
In the case of CMB data alone, we obtain 
$n_{\rm s} = 0.970^{+0.033(0.110)}_{-0.033(0.052)}$ where the errors 
correspond to  68\% (95\%), 
fully consistent with $n_{\rm s}=1$. However, with the
smaller errors afforded by combining the CMB data with the
2dFGRS $P(k)$ data, we obtain
$n_{\rm s} = 0.954^{+0.023(0.054)}_{-0.023(0.040)}$.
This measurement of the scalar perturbation spectral 
index is consistent with scale invariant value 
$n_{\rm s}=1$ at the $95\%$ level. Any detection of a deviation from 
scale invariance would have strong implications for the inflationary paradigm,
and we discuss this result in more detail below in Section~5.2.

\subsection{Six parameters plus the mass fraction of massive neutrinos}

Massive neutrinos were ruled out a generation ago as the sole 
constituent of the dark matter, on the basis of N-body simulations 
of the formation of large scale structure in hot dark matter 
universes (Frenk, White \& Davis 1983).  
However, interest in massive neutrinos has been resurrected recently 
with the resolution of the solar neutrino problem and the advent of 
precision measurements of the galaxy power spectrum. 
The detection of other flavours of neutrino in addition to 
the electron neutrino in the flux of neutrinos from the Sun 
suggests that neutrinos can oscillate between flavours (Ahmad et~al. 2001).  
This in turn implies that the three known types of neutrino have a 
non-zero mass, 
although measurements of the degree of flavour mixing set limits 
on the mass-squared differences between the neutrino 
flavours rather than on their absolute masses. 
The most extreme (and perhaps most plausible) case is where
the lightest mass eigenvalue is negligibly small, in which case
the sum of neutrino masses is dominated by the heaviest
eigenvalue: $\sum{m_{\nu}} \simeq m_3 \simeq 0.045$~eV 
(for a recent review see Barger et~al. 2003).
The only way in which $\sum{m_{\nu}}$ can greatly exceed this
figure is if the mass hierarchy is almost degenerate;
we therefore assume three species of equal mass in what follows.
Absolute measurements of neutrino mass can be obtained from 
Tritium beta decay experiments. At present,  
such experiments provide a limit on the sum of the neutrino masses of  
$\sum{m_{\nu}}< 6.6$~eV at the 2$\sigma$ level (Weinheimer 2002).

\begin{figure}
\includegraphics[width=0.47\textwidth]{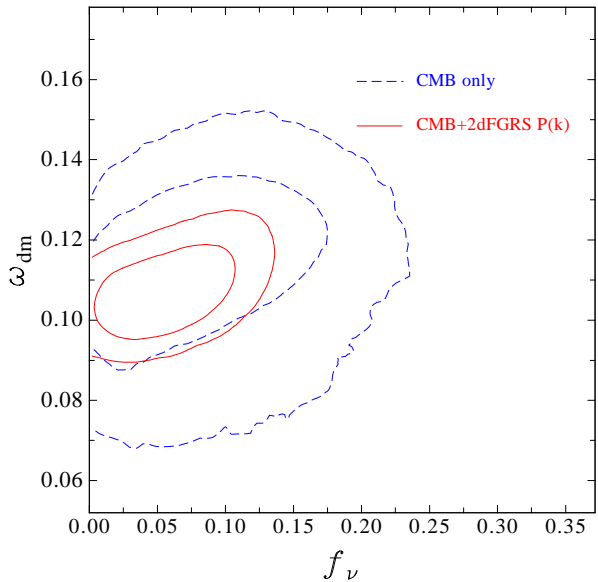}
\caption{
The marginalized posterior likelihood in the $f_{\nu}-\omega_{\rm dm}$ 
plane for the basic-six+$f_{\nu}$ parameter set. 
The dashed lines show the 68\% and 95\% contours obtained 
in the CMB only case. The solid contours 
show the corresponding results obtained in 
the CMB plus 2dFGRS $P(k)$ case. 
}
\label{fig:b6f-f-wdm}
\end{figure}

\begin{figure}
\includegraphics[width=0.47\textwidth]{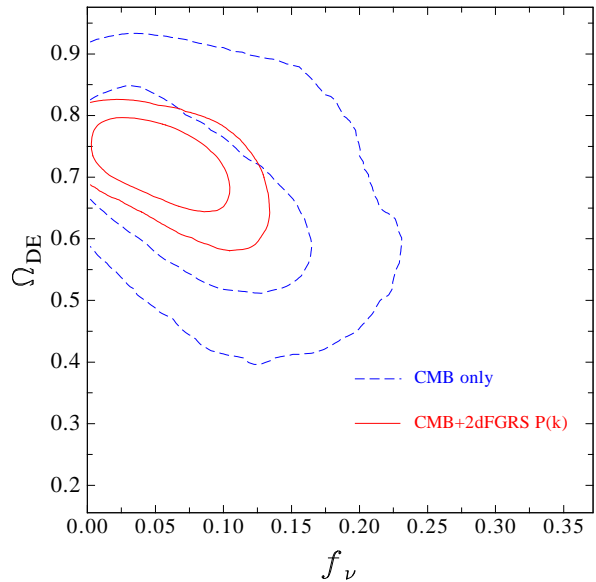}
\caption{
The marginalized posterior likelihood in the $f_{\nu}-\Omega_{\rm DE}$ 
plane for the basic-six+$f_{\nu}$ parameter set. The dashed lines 
show the 68\% and 95\% contours obtained in the CMB only case. 
The solid contours show the parameter constraints 
obtained for combined CMB and 2dFGRS $P(k)$ datasets. 
}
\label{fig:b6f-f-oml}
\end{figure}

Currently, the most competitive limits on neutrino masses are obtained 
through the comparison of CMB and LSS data with theoretical models 
(Hu et~al. 1998; Elgaroy et~al. 2002; Hannestad 2002).   
In the early universe, when neutrinos were still relativistic, they 
free-streamed out of density perturbations, damping 
overdensities in the baryons and cold dark matter. This smearing effect 
stops once neutrinos become non-relativistic; in this case, free-streaming 
only suppresses power on scales smaller than the horizon at this epoch, 
which depends on neutrino mass. 

The CMB temperature power spectrum is only weakly dependent on 
the neutrino mass fraction, $f_{\nu}$, since at the epoch of last scattering 
neutrinos with eV masses behave in a similar fashion to cold dark 
matter. So, CMB data alone do a poor job of constraining the neutrino 
mass fraction. Moreover, the response of the CMB power spectrum to variations 
in $f_{\nu }$ is limited to the higher multipoles ($\ell \ge 700$)
and so the first year WMAP data alone cannot give good 
constraints on this quantity (see the results of Tegmark et~al. 2004b). 
Our constraints in the CMB only case arise mainly due to data other than 
WMAP which probe smaller angular scales and therefore higher multipoles. 
On the other hand, the impact of massive neutrinos on the shape of the 
matter power spectrum is much more pronounced. The combination of CMB 
data with a measurement of the mass power spectrum can therefore give 
a much tighter constraint on the mass fraction of neutrinos; the shape 
of $P(k)$ constrains the value of $f_{\nu }$ while the CMB data sets 
the values of the parameters that are degenerate with $f_{\nu}$.

Using CMB data only, we find $f_{\nu}<0.182$ at 
$95\%$. When the 2dFGRS $P(k)$ is included 
this becomes $f_{\nu}<0.105$ at $95\%$. 
Our results can be converted into constraints on the 
sum of the three neutrino masses using $\sum {m_{\nu }} = \omega_{\rm dm}
f_{\nu}\,94.4\,\mathrm{eV}$ (assuming standard freezeout and that neutrinos 
are Majorana particles) to obtain the following limits: 
$\sum {m_{\nu}}<2.09\,$~eV at $95\%$ in the CMB only case,  
$\sum {m_{\nu }}<1.16 \,\rm{eV}$ at $95\%$ for CMB data 
plus the 2dFGRS $P(k)$.

Elgaroy et~al. (2002) used the Percival et~al. (2001) measurement 
of the 2dFGRS power spectrum to constrain the neutrino mass 
and found $\sum m_{\nu} < 2.2\,$~eV (95\%), assuming $n_{s}=1$ and a 
restrictive prior on $\Omega_{\rm m}$. Our results also represent a 
substantial improvement over those reported by Tegmark et~al. (2004b), 
who combined the first year WMAP data with the SDSS power spectrum to 
constrain a similar set of parameters to those we consider 
and found a $95\%$ limit of $\sum {m_{\nu }}\leq 1.7$~eV. 
Our results for $f_{\nu}$ provide an important illustration of the 
need to augment the WMAP data, which is the most accurate available 
for $\ell\le 600$, with measurements conducted at higher angular resolution, 
allowing significant improvements in the constraints attainable on 
certain parameters. 

It is possible to obtain a stronger limit from CMB+LSS studies
if amplitude information is also used: a neutrino fraction
reduces the overall growth rate as well as changing the 
shape of the matter power spectrum. This constraint was used
in the year-1 WMAP analysis, and was important in reaching the
tight constraint of $\sum m_{\nu} < 0.7$~eV
(Spergel et ~al. 2003; Verde et~al. 2003). This analysis
required the use of the 2dFGRS bispectrum in addition to
$P(k)$ (Verde et~al. 2002; for a determination with the 
final 2dFGRS see Gazta\~{n}aga et~al. 2005); we have 
preferred not to use this information at
the present time, since it has not been subject to the same
degree of detailed simulation as $P(k)$.
The limit on the neutrino mass can also be tightened
if a measurement of the linear theory matter power spectrum is 
available at higher wavenumbers than can be probed with the galaxy 
power spectrum. Seljak et al. (2005) used the power spectrum of the 
Ly-$\alpha$ forest and the SDSS $P(k)$, with a prior on the optical depth 
to last scattering of $\tau < 0.3$ (see later), to 
obtain $\sum {m_{\nu }} < 0.42$~eV. The extraction of the linear theory 
power spectrum of matter fluctuations from the Lyman-$\alpha$ forest remains 
controversial, so we do not address the use of this dataset here 
(Croft et~al. 2002; Gnedin \& Hamilton 2002; McDonald et~al. 2005). 

The only work to have reported a measurement of a non-zero neutrino mass
rather than an upper limit is Allen et~al. (2003). 
These authors combined galaxy cluster data with CMB data and an earlier 
version of the 2dFGRS $P(k)$ measured by Percival et~al. (2001).
The cluster data used by these authors was the gas fraction and the 
X-ray luminosity function; both quantities are much more difficult to model 
than the CMB and LSS data that we consider here. 
Although their results show a stronger signal upon the inclusion of 
the galaxy cluster data, there is still the suggestion of a non-zero 
neutrino mass fraction even with the CMB and 2dFGRS $P(k)$ data alone, 
showing that this conclusion is not due exclusively to the use of 
the X-ray data. The parameter space explored by Allen et~al. differs 
from the one considered in this section, since it includes tensor modes. 
The tensor modes contribute 
to the low multipole part of the CMB spectrum and their inclusion can drive 
down the amplitude of the scalar perturbations on these scales. 
This in turn can lead to an increase in the recovered value 
of the scalar spectral index, $n_{\rm s}$, with the consequence 
that $f_{\nu}$ increases to compensate, thus maintaining the power 
in the mass distribution at high $k$. 
This degeneracy in the $f_{\nu}-r$ plane produces a higher 
one-dimensional marginalized constraint on the neutrino mass 
fraction.

Fig.~\ref{fig:b6f-f-wdm} and Fig.~\ref{fig:b6f-f-oml} show the impact of 
including the 2dFGRS $P(k)$ data on the $f_{\nu}-\omega_{\rm dm}$ 
and $f_{\nu}-\Omega_{\rm DE}$ constraints. In the CMB only case, 
the incorporation of $f_{\nu}$ into the parameter space causes  
the uncertainty in all the parameters to grow. This is particularly 
noticeable for $\Omega_{\rm DE}$, for which the errors are twice 
as big as they were for the basic-six parameter set with $f_{\nu}=0$. 
When the 2dFGRS power spectrum is added to the analysis, the allowed 
ranges of these parameters are dramatically reduced, with particularly 
tight constraints resulting on $\omega_{\rm dm}$ and $\Omega_{\rm DE}$; 
this clearly demonstrates the importance of including LSS data to 
obtain precise constraints on these parameters.

\subsection{Six parameters plus the curvature of the universe: non-flat models}
\label{ssec:omegak}

\begin{figure}
\includegraphics[width=0.47\textwidth]{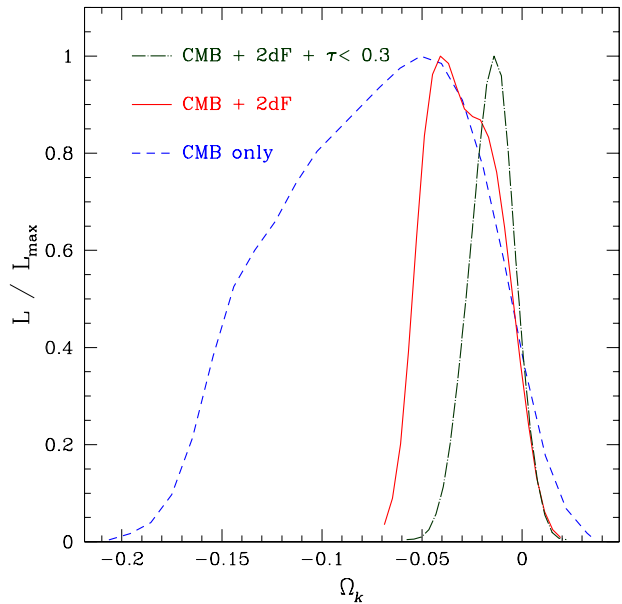}
\caption{
The one dimensional marginalized posterior likelihood for $\Omega_{k}$ 
for CMB data only (dashed line), CMB plus 2dFGRS $P(k)$ (solid line), 
and CMB plus 2dFGRS $P(k)$, with a prior on the optical depth 
of $\tau<0.3$ (dot-dashed line). 
Closed models have $\Omega_{k}<0$.
}
\label{b6wplot2}
\end{figure}

\begin{figure}
\includegraphics[width=0.47\textwidth]{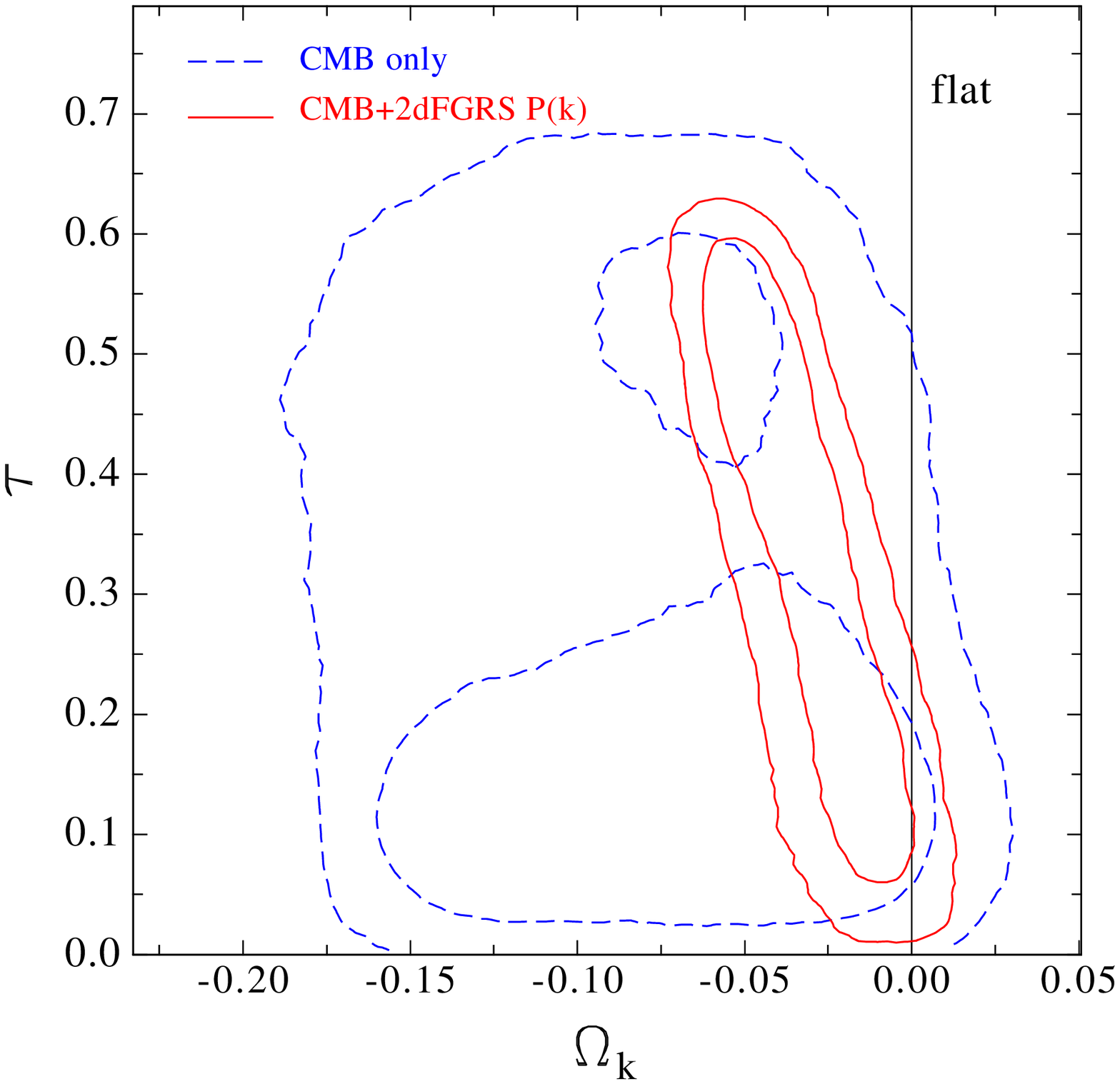}
\caption{
The marginalized posterior likelihood in the $\Omega_{k}-\tau$ plane for 
the basic-six  plus $\Omega_{k}$ parameter set. The dashed lines show 
the 68\% and 95\% contours obtained in the CMB only case. The 
solid contours correspond to the constraints obtained in the CMB plus 
2dFGRS $P(k)$ case. 
}
\label{b6omkplot3}
\end{figure}

\begin{figure}
\includegraphics[width=0.47\textwidth]{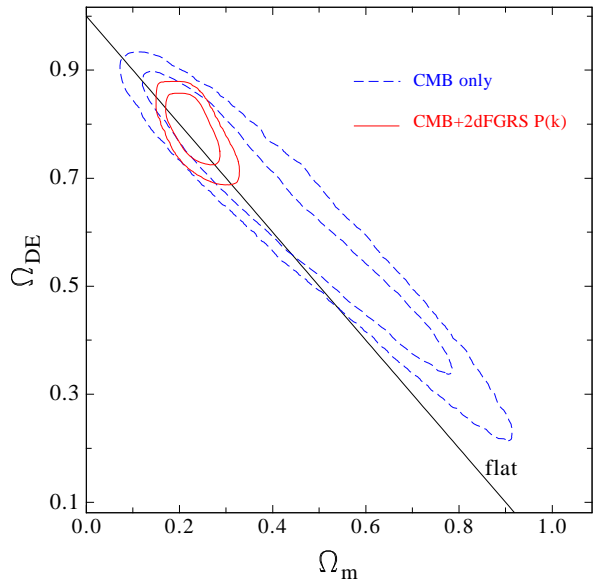}
\caption{
The marginalized posterior likelihood in the 
$\Omega_{\rm m}-\Omega_{\rm DE}$ plane for the basic-six plus 
$\Omega_{k}$ parameter set. The dashed lines show the 68\% and 
95\% contours obtained in the CMB only case. The solid contours 
correspond to constraints in the CMB plus 2dFGRS $P(k)$ case. 
}
\label{b6omkplot1}
\end{figure}

There is a strong theoretical prejudice that we live in a flat universe 
with $\Omega_{k}=0$. 
The first detections of the acoustic peaks in the CMB temperature power 
spectrum, the location of which is a measure of the geometry of the 
universe, showed that the universe is {\it close} to being flat (de Bernardis 
et~al. 2000). These results 
served to reinforce the prejudice that the curvature of the 
universe must be \textit{exactly} zero -- and it is true that, to date, no 
work has found any strong indication of a significant deviation from 
$\Omega_{k}=0$. However, as the flatness of the universe is one of 
the most important predictions of inflationary models, this assumption 
must be properly tested against new datasets. We must bear in mind, when 
comparing values reported for cosmological parameters, that many works 
simply assume $\Omega_{k}=0$. Other parameters, for example the 
scalar spectral index, are sensitive to the prior assumed for $\Omega_{k}$.

We plot the marginalized likelihood function for $\Omega_{k}$ in 
Fig.~\ref{b6wplot2}, for different datasets. The dashed curve shows the 
results for the CMB data alone, reminding us of the well-known
(but frequently forgotten) result
that the CMB data alone do not require a flat universe.  
Even though values of $\Omega_{k}>0$ (open models) are practically 
ruled out, a wide range of closed models is allowed, with the best fitting 
value given by $\Omega_{k}=-0.074_{-0.052\,(0.084)}^{+0.049\,(0.076)}$ at 68\% 
(95\%) confidence. 
The solid line in Fig.~\ref{b6wplot2} shows how incorporating the 2dFGRS 
power spectrum helps to tighten the constraints on $\Omega_{k}$. The 
addition of power spectrum information helps to break the geometrical
degeneracy between $\Omega_{\rm m}$ and $\Omega_{\rm DE}$ 
(see Fig.~\ref{b6omkplot1} and the final paragraph of this subsection). 
This is one of the most important effects of the incorporation of 
LSS information into the analysis. In the CMB plus 2dFGRS $P(k)$ case, 
we get $\Omega_{k}=-0.029^{+0.018\,(0.032)}_{-0.018\,(0.028)}$. 

It is particularly important to note the effect that the prior on the 
optical depth to the last scattering surface, $\tau$, has on the inferred value 
of the curvature of the universe. Fig.~\ref{b6omkplot3} shows the 
constraints in the $\Omega_{k}-\tau $ plane. The addition of the 
2dFGRS power spectrum shrinks the allowed region by tightening up the 
constraints on $\Omega_{k}$, but the resulting likelihood contours show a 
clear degeneracy between the two parameters, with the high values of 
$\tau$ preferring more negative values of $\Omega_{k}$. This degeneracy is 
responsible for the broad error bars on these parameters. 
If one adopts a restrictive prior on the optical depth of $\tau <0.3$, 
as recommended by the WMAP team based on the lack of a large
signal in polarization autocorrelation, then 
the results for $\Omega_{k}$ are more in line with those in the 
literature, as shown by the dot-dashed line in Fig.~\ref{b6wplot2}. 
In this case we find 
$\Omega_{k}=-0.015_{-0.011\,(0.020)}^{+0.011\,(0.023)}$ 
for the combined CMB plus 2dFGRS datasets.
We shall return to the issue of the choice 
of prior for the optical depth in Section~\ref{ssec:ptau}. 

Finally, we highlight the constraints on the densities of dark matter 
and dark energy obtained when the assumption of a flat universe is 
dropped. Fig.~\ref{b6omkplot1} shows the results for the case of CMB data 
alone (dashed lines) and for CMB data plus the 2dFGRS $P(k)$ (solid lines). 
As we have seen in several previous examples, there is a dramatic improvement 
in the quality of the constraints on these parameters once the galaxy 
clustering  
data is incorporated into the analysis. There is compelling evidence for a 
dark energy component in the universe. 

\subsection{Six parameters plus the dark energy equation of state}

Over the past decade, mounting evidence has been presented for the  
accelerating expansion of the Universe, based on the interpretation 
of the Hubble diagram of Type-1a supernovae (Perlmutter et~al. 1999; 
Riess et~al 2004). 
Independent support for the presence of a dynamically dominant, 
negative pressure component in the energy-density budget of the Universe 
has also come from fitting cosmological models to CMB and LSS datasets 
(see Section~\ref{ssec:omegak} and Efstathiou et~al. 2002).   
Although we can infer the presence of this component, dubbed dark energy, 
we know practically nothing about its nature. A plethora of theoretical 
models have been proposed for the dark energy (e.g. see the review by 
Sahni 2004). 
One of the key properties of the dark energy which can be used to 
pare down the market of possible models is the equation of state of 
the dark energy, that is the ratio of its pressure to density, $w_{\rm DE}$. 

\begin{figure}
\includegraphics[width=0.47\textwidth]{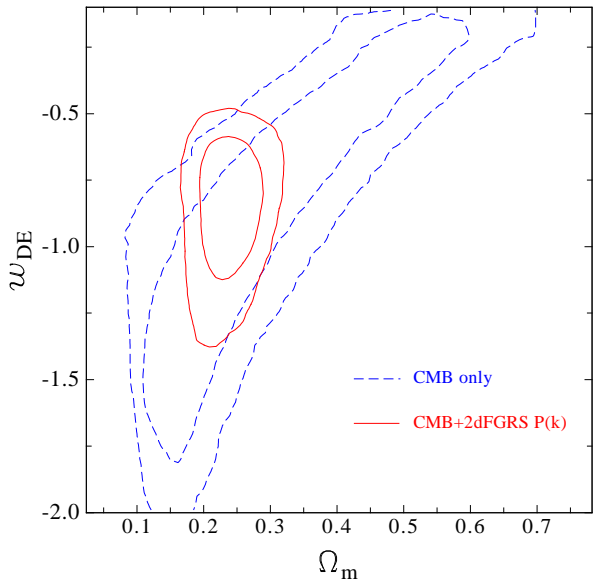}
\caption{
The marginalized posterior likelihood in the $\Omega_{\rm m}-w_{\rm DE}$ 
plane for the basic six plus $w_{\rm DE}$ parameter set. 
The dashed lines show the 68\% and 95\% contours obtained 
in the CMB only case. The solid contours show the corresponding 
constraints obtained in the CMB+2dFGRS $P(k)$ case. 
}
\label{fig:b6w-omm-w}
\end{figure}

Until now we have assumed that the dark energy component corresponds to 
the cosmological constant,  with a fixed equation of state specified by 
$w_{\rm DE}=-1$. However, this is only one manifestation of the many possible 
forms that the dark energy could take. Any component with an equation of state 
$w_{\rm DE}<-1/3$ will result in an accelerating rate of expansion today. 
In this section, we explore dark energy models with a constant equation of 
state, allowing for variations in the redshift independent value of 
$w_{\rm DE}$. We also consider models with $w_{\rm DE}<-1$, 
sometimes referred to as `phantom
energy'. 

Fig.~\ref{fig:b6w-omm-w} shows the marginalized constraints in the 
$w_{\rm DE}-\Omega_{\rm m}$ plane. In the CMB only case, we find  
$w_{\rm DE}=-0.93_{-0.47}^{+0.49}$, consistent with a cosmological 
constant. When the 2dFGRS power spectrum is included in the analysis, 
the preferred value increases somewhat to $w_{\rm DE}=-0.85_{-0.17}^{+0.18}$. 
If we also include the supernova type Ia data from 
Riess et~al. (2004), our result scarcely changes, with 
$w_{\rm DE}=-0.87_{-0.12}^{+0.12}$. 
Phantom energy models are permitted in the case of CMB data only, 
with the $95\%$ limit on the equation of state of $w_{\rm DE}>-1.66$. 
However, once the 2dFGRS $P(k)$ and supernovae type-Ia data are included,  
the allowed region shrinks to a smaller zone 
with $w_{\rm DE}>-1.19$ at $95\%$, 
showing that phantom energy models are disfavoured by the currently available 
data. 
These results show that the data prefers lower values of $w_{\rm DE}$ 
than suggested by previous work using the SDSS power spectrum (MacTavish  
et~al. 2005). Our results are consistent with the dark energy taking 
the form of a cosmological constant.  
We will discuss this point further in Section~\ref{sec:sdss}.

\begin{figure}
\includegraphics[width=0.47\textwidth]{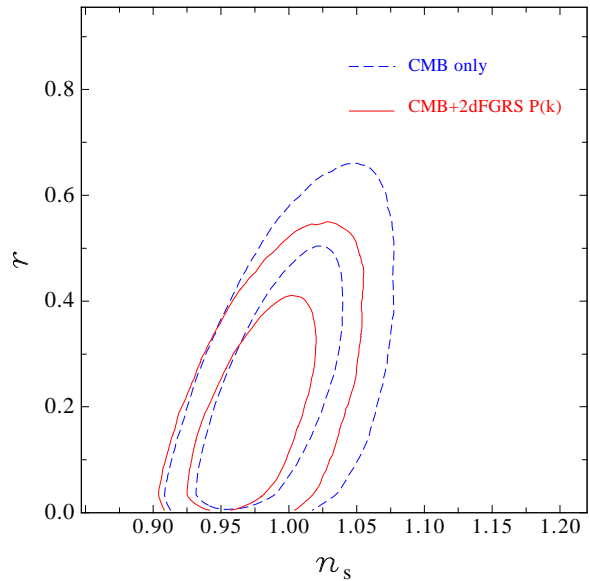}
\caption{
The marginalized posterior likelihood in the $n_{\rm s}-r$ plane for 
the basic-six plus $r$ parameter set. The dashed lines show the 68\% and 
95\% contours obtained in the CMB only case. The solid contours 
show the corresponding results in the CMB plus 2dFGRS $P(k)$ case. 
}
\label{fig:b6r-ns-r}
\end{figure}

\subsection{Six parameters plus non-zero tensor modes}

We now add the ratio of the amplitude tensor to scalar 
perturbations, $r$, to the basic-six parameter set. 
This case is an important one to consider as tensor modes 
are predicted to be present in many inflationary models. 
Moreover, as we shall see, several cosmological parameters 
are degenerate with $r$ and in the literature tensor modes have often 
been ignored when presenting constraints on these parameters. 

\begin{figure}
\includegraphics[width=0.47\textwidth]{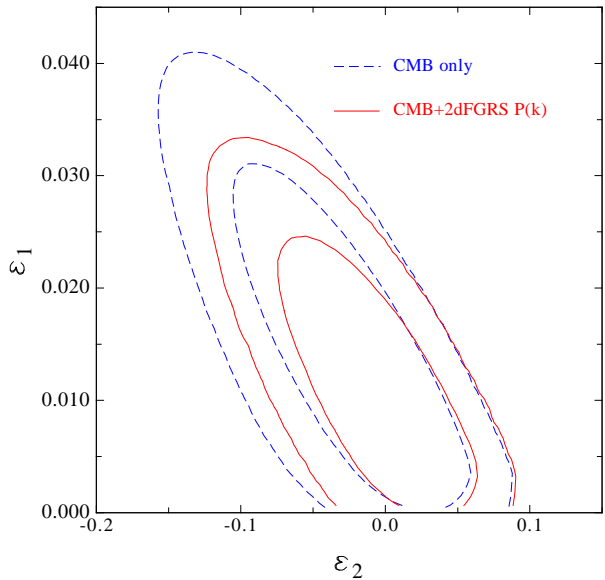}
\caption{
The marginalized posterior likelihood in the $\epsilon_1-\epsilon_2$ plane for 
the basic-six plus $r$ parameter set. The dashed lines show the 68\% and 
95\% contours obtained in the CMB only case. The solid contours 
correspond to the results obtained in the CMB plus 2dFGRS $P(k)$ case. 
}
\label{fig:b6r-e1-e2}
\end{figure}

\begin{figure}
\includegraphics[width=0.47\textwidth]{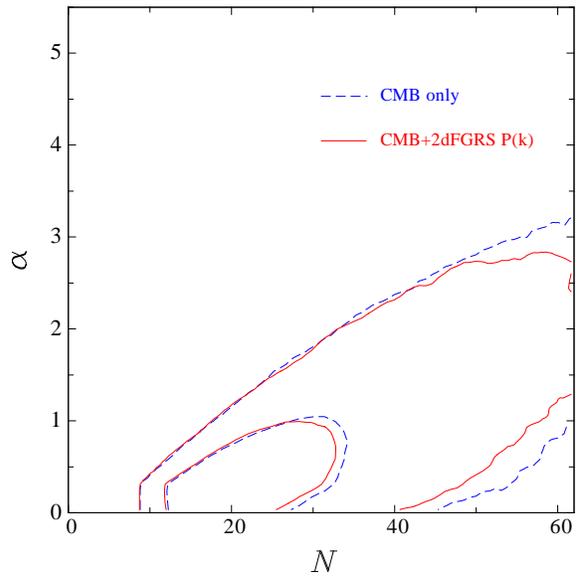}
\caption{
The marginalized posterior likelihood in the $\alpha - N$ plane for 
the basic-six plus $r$ parameter set. The dashed lines show the 68\% and 
95\% contours obtained in the CMB only case. The solid contours 
show the results in the CMB plus 2dFGRS $P(k)$ case. 
}
\label{fig:b6r-alpha-n}
\end{figure}

The constraints imposed on $r$ by CMB information alone are 
$r<0.52$ at $95\%$. Including the 2dFGRS 
$P(k)$ data reduces the importance of tensors slightly, 
yielding $r<0.41$ at $95\%$. 
Fig.~\ref{fig:b6r-ns-r} shows the two-dimensional marginalized likelihood 
contours in the $n_{\rm s}-r$ plane for the cases of CMB data only (dashed 
lines) and CMB plus the 2dFGRS $P(k)$ (solid lines). 
Tensor modes contribute to the CMB temperature power spectrum only 
on large angular scales, leading to a reduction in the scalar perturbations 
on these scales to match the observations. In order to maintain the 
amplitude of scalar perturbations on smaller angular scales, an increase 
in the scalar spectral index, $n_{\rm s}$, is required. 
This degeneracy results in a broader allowed range for $n_{\rm s}$ 
than in the case where only scalar modes are considered. 

The constraints on $r$ and $n_{\rm s}$ can be translated into 
the horizon flow parameters, $\epsilon_1$ and $\epsilon_2$, using 
the relations given by Mukhanov et~al. (1992):
\begin{eqnarray}
 1 - n_{\rm s} &=& 2\epsilon_1 + \epsilon_2 \\
 r &=& 16\epsilon_1. 
 \label{eq:n-r}
\end{eqnarray}
The horizon flow parameters are related to derivatives of the Hubble 
parameter during inflation (Schwarz et~al. 2001).
Leach \& Liddle (2003a) give equations relating the horizon flow 
parameters to the derivatives of the inflaton potential and discuss 
the motivation for the truncation of the slow-roll expansion 
after $\epsilon_2$.   
The constraints on the horizon flow 
parameters are shown in Fig.~\ref{fig:b6r-e1-e2}. 
The degeneracy between $r$ and $n_s$ translates into a degeneracy in 
$\epsilon_1$ and $\epsilon_2$. 

If we restrict our attention to monomial inflation, i.e. potentials  
of the form $V \propto \phi^{\alpha}$, then the horizon flow parameters 
can be related to the power law index, $\alpha$, and the number 
of $e$-folds of inflation for the scale considered, $N$, 
by the simple relations 
(Leach \& Liddle 2003b):
\begin{eqnarray}
 \epsilon_2 & = & \frac{4\epsilon_1}{\alpha} \\
 N & = & \frac{\alpha}{4}\left(\frac{1}{\epsilon_1}-1\right). 
 \label{eq:alfa-n}
\end{eqnarray}
To obtain constraints on these new parameters, we have 
translated our results for $\epsilon_1$ and $\epsilon_2$ into the 
$\alpha - N$ plane. 
In doing so, we have restricted our attention to the region 
where $\epsilon_2>0$, following Liddle \& Leach (2003b), who 
argue that this part of the horizon flow parameter space 
contains the most likely models in which inflation will end 
naturally with a violation of the slow-roll approximation. 
Our results are plotted in Fig.~\ref{fig:b6r-alpha-n}. 
We find that $\alpha < 2.33$ at $95\%$ for CMB data alone and 
$\alpha < 2.27$ (95\%) for CMB plus the 2dFGRS $P(k)$. 
To obtain this result, we have followed Seljak et~al. (2005) 
and take into account the maximum number of e-folds, 
$N_{\rm max}=60$,  of slow-roll inflation experienced at 
the pivot scale $k=0.05\,\mathrm{{Mpc^{-1}}}$, thus further 
restricting the second horizon flow parameter, $\epsilon_{2} > 0.0167$.
(Note that Leach \& Liddle 2003b use a different pivot scale to ours.) 
Our constraint on $\alpha$ implies that the $\lambda\phi^{4}$ inflation 
model is ruled out. 
This is the first time that the CMB data alone have 
been of sufficient quality to completely reject this model. 
Seljak et~al. (2005) reached a similar conclusion using different datasets: 
the WMAP data, the SDSS galaxy power spectrum and either the power 
spectrum of the Ly-$\alpha$ forest or the amplitude of the matter 
power spectrum, as inferred from the bias of SDSS galaxies. 

We now turn our attention to power-law inflationary models, in 
which the scale factor of the universe grows with time as   
$a \propto t^{p}$ with $p>1$. In such models, the horizon flow 
parameters are simply given by
\begin{eqnarray}
 \epsilon_1 &=& \frac{1}{p} \\
 \epsilon_{\rm i} &=& 0 \,\,\, i \ge 2 . 
 \label{eq:power-law}
\end{eqnarray}
Substitution of these relations in Eq.~\ref{eq:n-r} 
gives a relation between $n_{\rm s}$ and $r$: 
\begin{equation}
r=8(1-n_{\rm s}).  \label{eq:rfijo}
\end{equation}
To analyse this kind of model, we ran a new set of chains fixing the
tensor to scalar ratio using Eq.~\ref{eq:rfijo}. 
In this case we get $n_s=0.978^{+0.010}_{-0.010}$ for the CMB data alone 
and $n_s=0.9762^{+0.0094}_{-0.0092}$ in the CMB plus 2dFGRS case. 
The constraints on $r$ are also tighter, with $r<0.31$ at $95\%$. 
We note that the best fitting values for the horizon flow
parameters of $\epsilon_1=0.0123^{+0.0080}_{-0.0082}$ (corresponding 
to $p=81^{+163}_{-32}$) and $\epsilon_2=-0.004^{+0.040}_{-0.040}$ 
are in complete agreement with the power law inflation picture.

\section{The role of priors}

It is often claimed that we have entered an era of precision cosmology, in
which the values of the cosmological parameters are known with high accuracy.
The CMB measurements alone go a long way towards realizing this ideal, 
but ultimately fall short due to the presence of well 
known degeneracies between the cosmological parameters 
(Efstathiou \& Bond 1999). 
Some of these degeneracies can be broken with the incorporation of other 
information into the analysis (such as, for example, LSS, SN Ia or the power 
spectrum of the Ly-$\alpha $ forest). 
However, many degeneracies remain even after the addition of these datasets.
Another way to break degeneracies is by imposing priors on parameters, which 
can have implications for the derived parameter constraints (see, for 
example, Bridle et~al. 2003a). 
In this section we revisit the constraints obtained for the different 
parameter sets and priors and assess which of our results are the most robust. 

\subsection{The baryon density}

One of the most important achievements of modern cosmology is the agreement
between the value of the physical density 
of baryons determined from CMB data and that  
inferred from Big Bang Nucleosynthesis (BBN) arguments and distant quasar 
absorption spectra.  
In the present analysis, we obtain a value for the baryon density 
of $\omega_{\rm b} = 0.0229^{+0.0012}_{-0.0013}$ from 
the CMB data alone that is consistent with the latest 
constraint from BBN: $\omega_{\rm b} = 0.022\pm 0.002$ (Cuoco et~al. 2004). 
This agreement is reinforced when the 2dFGRS $P(k)$ is added to the analysis, 
with $\omega_{\rm b} = 0.0225^{+0.0010}_{-0.0010}$. 
The variation in the value of $\omega_{\rm b}$ obtained between the different 
parameter sets and priors that we have analysed is smaller than 
the 1$\sigma$ error bars, showing the robustness of this result.
This level of agreement is all the more remarkable when one considers the 
quite different epochs to which the various datasets relate: BBN is a theory 
that describes processes occurring in the very early universe, just a few 
minutes after the Big Bang, while the CMB maps the universe 
as it was a few hundred thousand years after the Big Bang, and the 
galaxy power spectrum refers to the present day universe, over 13 billion 
years later. The fact that we can tell a coherent story over such a huge 
baseline in time and physical conditions provides an impressive verification of 
our cosmological model.

\subsection{The dark matter density}

A scan across the third rows of Tables 2 and 3 shows that the 
value of $\omega_{\rm dm}$ is largely insensitive to the priors applied  
to the other parameters. The one exception is when the flatness prior,  
$\Omega_{k}=0$, is relaxed, in which case we obtain a
smaller value for $\omega_{\rm dm}$ with larger errors. 
The constraints obtained on $\omega_{\rm dm}$ in the CMB only and 
CMB plus 2dFGRS $P(k)$ cases are fully consistent. 

\begin{figure}
\includegraphics[width=0.47\textwidth]{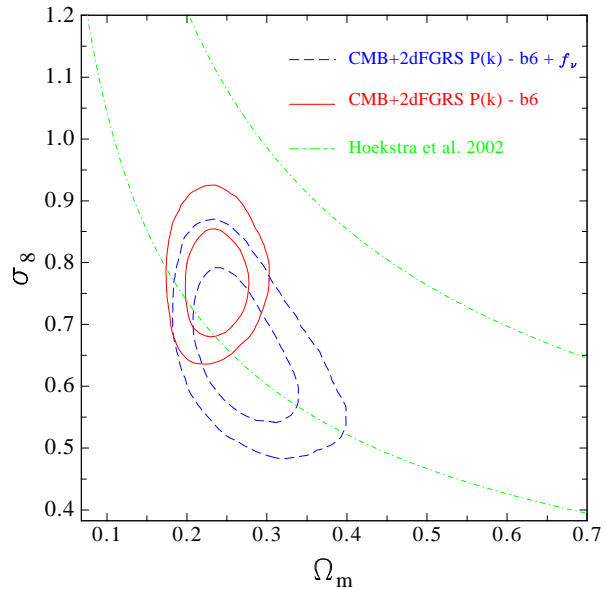}
\caption{
The marginalized posterior likelihood in the $\Omega_{\rm m}-\sigma_{8}$ plane 
obtained using CMB plus 2dFGRS information for different parameter sets. 
The solid lines correspond to the 68\% and 95\% contours 
obtained for the basic-six parameter set. 
The dashed lines correspond to the results obtained when the 
neutrino fraction $f_{\nu}$ is also allowed to vary. 
The dot-dashed lines show constraints from weak lensing measurements 
from Hoekstra et~al. (2002).
}
\label{fig:priors_16_17}
\end{figure}

\begin{figure}
\includegraphics[width=0.47\textwidth]{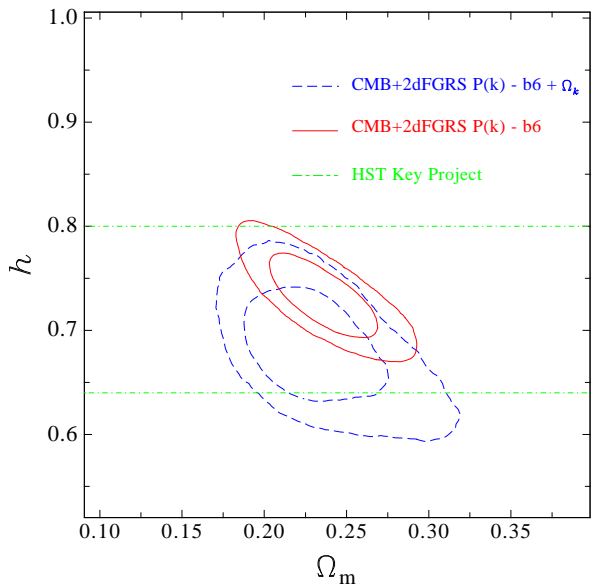}
\caption{
The marginalized posterior likelihood in the $\Omega_{\rm m}-h$ plane 
obtained using CMB plus 2dFGRS information for different parameter sets. 
The solid lines correspond to the 68\% and 95\% contours 
obtained for the basic-six parameter set. The dashed lines show the results 
obtained when non-flat models are considered (basic-six 
plus $\Omega_{k}$). 
The dot-dashed lines show the 1$\sigma$ constraint on $h$ 
from the HST Key Project (Freedman et~al. 2001).
}
\label{fig:priors_16_20}
\end{figure}

The implications of the value of $\omega_{\rm dm}$  for the matter density 
$\Omega_{\rm m}$ do, however, depend on the priors implemented. 
In the basic-six plus $f_{\nu }$ parameter set, for the CMB plus 2dFGRS 
case we find  $\Omega_{\rm m}=0.282\pm 0.040$, but it can be as low 
as $\Omega_{\rm m}=0.224 \pm 0.022$ for the basic-six plus $r$ parameter 
set. 
With the exception of the case of non-zero neutrino mass,
all our estimates of $\Omega_{\rm m}$ lie significantly
below the standard choice of 0.3.
Fig.~\ref{fig:priors_16_17} illustrates how the choice of parameter 
space affects the results obtained. The constraints in the 
$\Omega_{\rm m}-\sigma_{8}$ plane in the basic-six parameter set 
are tighter than those obtained when the neutrino fraction, $f_{\nu}$, 
is incorporated into the analysis; for the latter case, a bigger region 
with lower values of $\sigma_{8}$ is allowed by the data.
A similar situation can be seen in Fig.~\ref{fig:priors_16_20} for the 
$\Omega_{\rm m}-h$ plane. The values of $h$ preferred by the data are lower 
when non-flat models are considered in the analysis.
These discrepancies cause differences in the marginalized results 
obtained for these parameters.
This situation occurs in many other cases and in general the 
influence of the parameter set is non-negligible. For this 
reason, constraints on a given parameter should always be quoted 
together with the parameter space explored in the analysis.
Fig.~\ref{fig:priors_16_20} shows the 1$\sigma$ limits on the 
Hubble constant derived by the HST Key Project (Freedman et~al. 2001). 
These constraints on $h$ are substantially broader than those obtained 
from the CMB plus 2dFGRS power spectrum, showing that including 
the Key Project measurement as a prior would have little impact on 
our results. 

\subsection{The amplitude of fluctuations}

When constraining the values of the cosmological parameters, we only use
information from the shape of the galaxy power spectrum and not from its
amplitude. Therefore the constraints on the amplitude of density 
fluctuations come principally 
from the CMB data, with the LSS data playing an indirect role by tightening 
the constraints on parameters which yield degenerate predictions 
for the CMB.  
The recovered values of $\sigma_{8}$ range from 
$\sigma_{8}=0.678_{-0.072}^{+0.073}$ for the basic-six 
plus $f_{\nu}$ parameter space to $\sigma_{8}=0.817_{-0.079}^{+0.077}$ 
for non-flat models (basic-six plus $\Omega_{k}$). 
Adding more data sets, such as the X-ray cluster luminosity function, 
or other measurements of the amplitude of fluctuations may help to 
improve the constraint on $\sigma_{8}$, but the theoretical  
modelling of these observational datasets is less straightforward.
In Fig.~\ref{fig:priors_16_17} we compare our results with constraints 
from measurements of weak lensing from Hoekstra et~al. (2002). 
Henry (2004) used the temperature function of X-ray clusters to 
find $\sigma_{8}=0.66 \pm {0.16}$, in good agreement with the 
basic-six plus neutrino mass fraction model. 
Similar constraints have been obtained by other groups 
(Bacon et~al. 2003; Heymans et~al. 2005).

\subsection{The optical depth}
\label{ssec:ptau}

\begin{figure}
\includegraphics[width=0.47\textwidth]{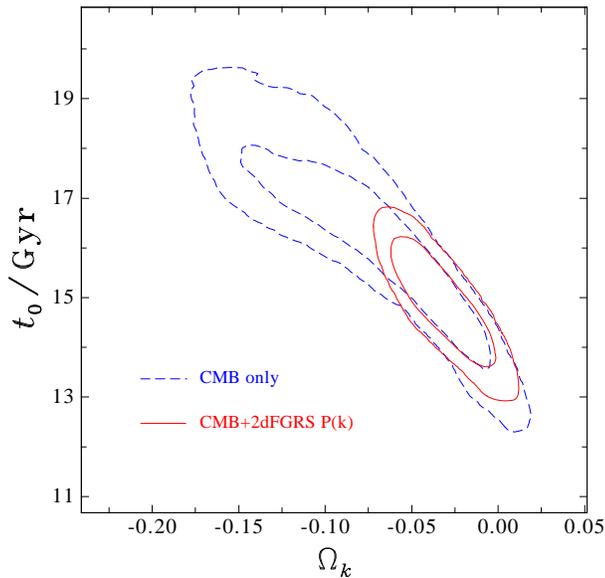}
\caption{
The marginalized posterior likelihood in the $\Omega_{k}-t_{0}$ plane for 
the basic-six plus $\Omega_{k}$ parameter set. The dashed lines 
show the 68\% and 95\% contours obtained in the CMB only case. 
The solid contours correspond to the constraints obtained in the 
CMB plus 2dFGRS case. 
}
\label{fig:prior-t0}
\end{figure}

\begin{figure}
\includegraphics[width=0.47\textwidth]{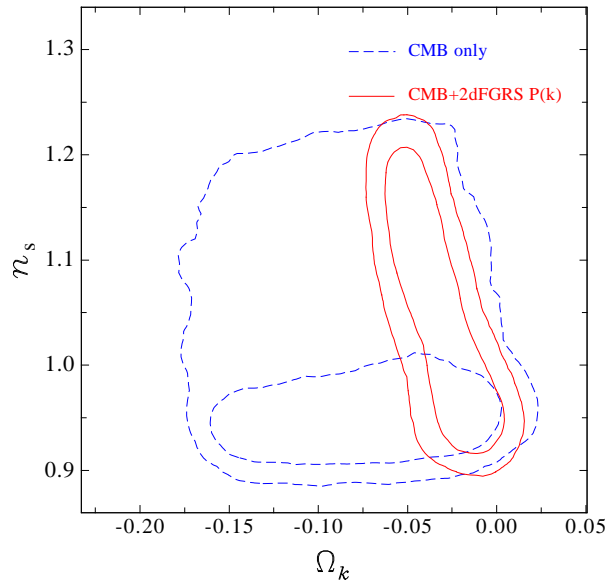}
\caption{
The marginalized posterior likelihood in the $\Omega_{k}-n_{\rm s}$ 
plane for the basic six plus $\Omega_{k}$ parameter set. The dashed 
lines show the 68\% and 95\% contours obtained in the CMB only 
case. The solid contours correspond to the constraints obtained in 
the CMB plus 2dFGRS  case. 
}
\label{fig:prior-omk}
\end{figure}

The optical depth to the last scattering surface has an important effect 
on nearly all other cosmological parameters. The signal for $\tau >0$ comes 
from the temperature-polarization cross-correlation on large angular scales. 
Intriguingly, Hansen et~al. (2004) performed a temperature-polarization 
analysis of the first year WMAP data for the northern and southern 
hemispheres separately and found that, whereas 
the northern hemisphere points to $\tau =0$, the southern hemisphere 
prefers a value of $\tau =0.24_{-0.07}^{+0.06}$, inconsistent 
with  $\tau =0$ at the 2$\sigma $ level, with the suggestion that 
the signal for $\tau >0$ may be due to foreground 
structures in the southern hemisphere. 

In their analysis of the first year WMAP data Spergel et~al. 
(2003) imposed a prior of $\tau <0.3$, justifying this by 
the need to avoid `unphysical' regions of parameter space. 
In Section~3.4 we demonstrated, as previously pointed out by 
Tegmark et~al. (2004b), the strong effect this prior has on 
our results. 
In particular, the $\tau <0.3$ prior is required to reconcile 
the constraints on $\Omega_{k}$ with the flatness prediction 
from inflation in the basic-six plus $\Omega_{k}$ parameter set, 
and to produce tight constraints on neutrino masses in the basic-six 
plus $f_{\nu}$ case. 
The situation should improve with the release of the second and 
subsequent years of data from WMAP, which will be
able to produce improved polarization maps.
In the meantime, the effect of this important 
prior must be borne in mind when interpreting the results from 
multi-parameter analyses.

\subsection{The flatness prior}

The prior of $\Omega_{k} = 0$ is widely used when constraining 
cosmological parameters. It is important to remember that, if the 
assumption of flatness is relaxed, the preferred value is actually 
$\Omega_{k}<0$ and only marginally consistent with $\Omega_{k} = 0$. 
The assumption of flatness has a major impact in the values of many 
cosmological parameters. 

The value obtained for the age of the universe, $t_{0}$, 
shows an important change when $\Omega_{k}$ is allowed to float, 
and is only marginally consistent with the values found for $\Omega_{k}=0$. 
Fig.~\ref{fig:prior-t0} shows the marginalized two-dimensional 
likelihood contours in the $\Omega_{k}-t_{0}$ plane. There is a 
clear degeneracy between these two parameters, with lower values of 
$\Omega_{k}$ preferring higher values of $t_{0}$; the incorporation 
of the 2dFGRS $P(k)$ data does not break this degeneracy completely. 
The same degeneracy can be seen in the $\Omega_{k}-h$ plane, 
which implies that a prior on the Hubble constant from the HST key project 
(Freedman et~al. 2001) may improve the situation, but even then the 
constraints on these parameters are less robust than in the flat case.

The scalar spectral index, $n_{\rm s}$, also merits special attention. 
In Section 3.2, we pointed out that the constraint on $n_{\rm s}$ in the 
basic-six model is only marginally consistent with $n_{\rm s}=1$; 
this spectrum is formally excluded at the $95\%$
confidence level. 
Fig.~\ref{fig:prior-omk} shows the marginalized two-dimensional likelihood 
contours in the $\Omega_{k}-n_{\rm s}$ plane. When CMB information alone 
is used, there is a wide allowed region that shrinks considerably when the 
2dFGRS power spectrum is included. In the latter case, there is a 
correlation between the parameters which makes the constraints on $n_{\rm s}$ 
much broader than those obtained for the special case of $\Omega_{k}=0$. 
Taking into account that the evidence for $n_{\rm s} < 1$ is weaker once 
more general parameter sets are considered (such as, for example, the 
basic-six plus $f_{\nu}$ set) and that the current data has a slight 
preference for closed models (even when the prior of $\tau<0.3$ is applied), 
we advocate caution before claiming a detection of a significant 
deviation from scale invariance.

\subsection{Tensor modes}

Another commonly applied prior is the assumption that tensor modes 
can be neglected. 
It is important to include the amplitude of tensor modes as a 
free parameter, not only because this has strong implications for 
inflationary models, but also because many other parameters are 
degenerate with the amplitude of tensors, resulting in the growth 
of the allowed regions for these parameters. The parameters that 
are most strongly influenced by the assumption about tensor modes 
are $n_{\rm s}$, $\Omega_{\rm m}$ and $h$. 

\section{Beyond the simplest model}

\subsection{How many parameters should float?}

We have shown that a model in which five parameters are allowed 
to vary gives a good description of the CMB and LSS datasets. 
We then went on to explore six and seven parameter sets, finding 
that, in some cases, the results obtained for certain parameters 
depended upon the choice of parameters varied. But are we justified 
in adding extra free parameters to our basic-five set? 

\begin{figure*}
\centering
\centerline{\includegraphics[width=0.95\textwidth]{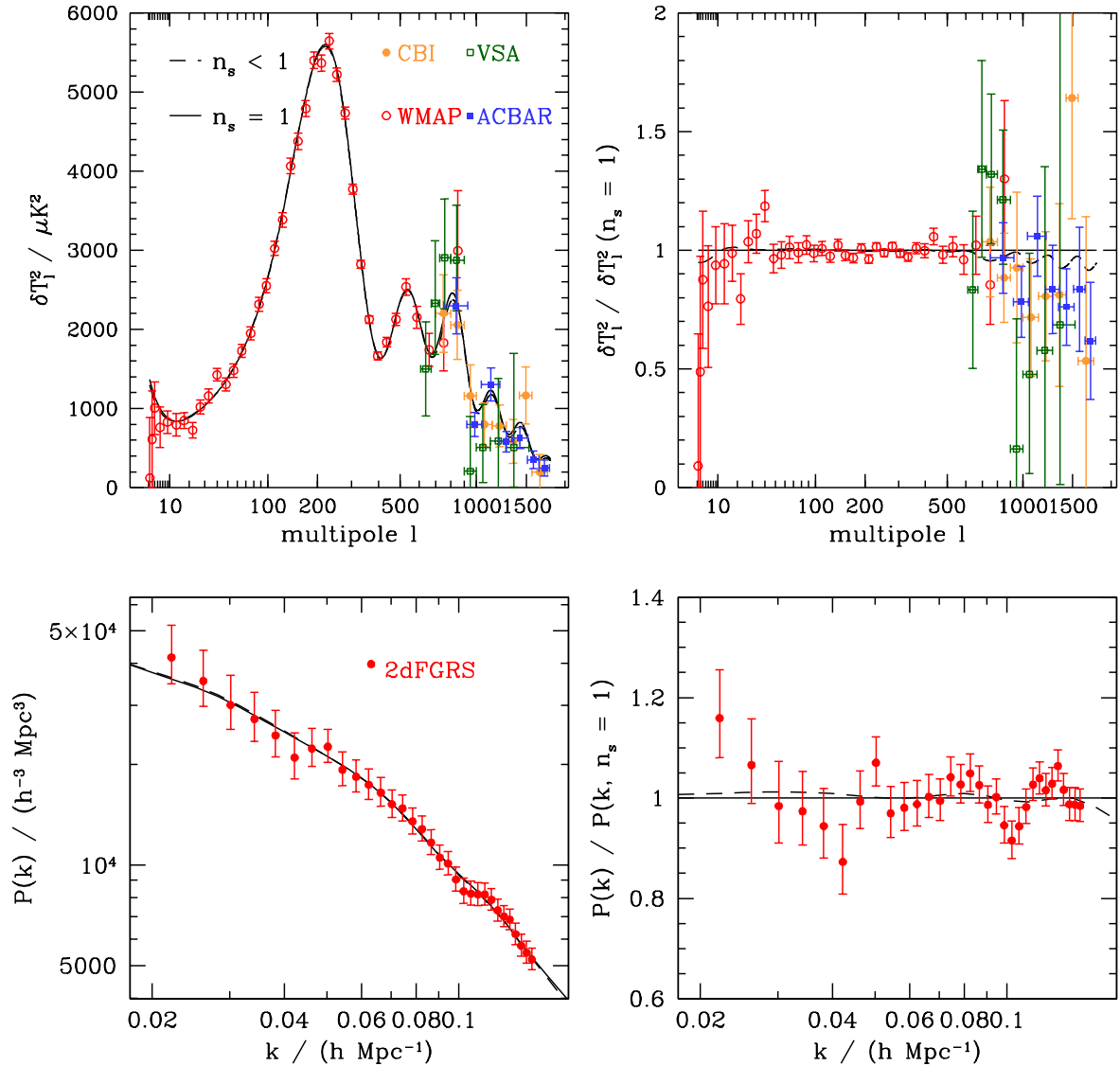}}
\caption{
Model fits to the CMB datasets (top panels) and 2dFGRS $P(k)$ (bottom panels) 
in the cases of the basic-five (solid line) and basic-six (dashed
line) models, with the best fitting parameter values listed in Table~3. 
The model $P(k)$ have been convolved with the window function of the 
2dFGRS from Cole et al. (2005). 
In the right-hand panels, the model curves and datapoints have been divided 
by the best fitting basic-five model to expand the y-axis. 
}
\label{fig:experiments}
\end{figure*}

The simplest way to make an objective assessment of different models 
is to establish whether or not they afford a better description of 
the data, which is usually done by computing likelihood ratios. However, 
it is important to compensate for the fact that adding extra parameters 
should necessarily improve the fit to the data. Liddle (2004) has 
advocated the use of two simple statistics that quantify the level of 
improvement in the description of cosmological datasets as new free parameters 
are added to the theoretical models. These statistics allow us to ascertain 
whether the addition of a new parameter is justified, i.e. does it 
produce a better than expected enhancement in the accuracy with which 
the data is reproduced? 
The statistics, the Akaike Information Criterion ({\tt AIC}; Akaike 1974) and 
the Bayesian Information Criterion ({\tt BIC}; Schwarz 1978) have a long 
track record of application in other branches of physics, but have largely 
been ignored in cosmology. 
The definitions of the two statistics are straightforward: 
\begin{eqnarray}
{\tt AIC}&=&-2\ln(\mathcal{L}) + 2N_{\rm par},  \label{eq:aic} \\
{\tt BIC}&=&-2\ln(\mathcal{L}) + N_{\rm par}\ln(N_{\rm data}),  \label{eq:bic}
\end{eqnarray}
where $\mathcal{L}$ is the maximum likelihood, $N_{\rm par}$ is 
the number of parameters varied in the model and $N_{\rm data}$ is the 
number of datapoints included in the analysis. 
The model that best describes the data with the most economical use of 
parameters is the one that minimizes these quantities. 
In both expressions, the first term favours models which provide better 
fits to the data, while the second penalizes large numbers of parameters. 
We note that, as the value of $\ln(N_{\rm data}) > 2 $ in our 
analysis, the {\tt BIC} actually gives a higher penalty to the number of 
free parameters than is the case for the {\tt AIC}. 

\begin{table}
\begin{center}
\caption{ The number of parameters, the likelihood and the values of the 
{\tt AIC} and {\tt BIC} statistics for the various models analysed in 
this paper. In all cases, $N_{\rm data}=1403$.}
\end{center}
\begin{center}
\begin{tabular}[t]{ccccc}
\hline\hline
Model              & $N_{\rm par}$  & $-2\ln(\mathcal{L})$ & {\tt AIC} & {\tt BIC} \\ \hline\hline
b5                 & 5          & 1495.8             & 1505.8 & 1532.1  \\ 
b6                 & 6          & 1492.1             & 1504.1 & 1535.6  \\ 
b6 + $f_{\nu}$     & 7          & 1491.3             & 1505.3 & 1542.0  \\ 
b6 + $\Omega_{k}$  & 7          & 1490.4             & 1504.4 & 1541.1  \\ 
b6 + $w_{\rm DE}$      & 7          & 1491.5             & 1505.5 & 1542.2  \\ 
b6 + $r$           & 7          & 1491.7             & 1505.7 & 1542.4  \\ 
\hline\hline
\end{tabular}
\end{center}
\label{tab:bic}
\end{table}

Table~4 provides a summary of the number of parameters, the likelihood 
and the values of the {\tt AIC} and {\tt BIC} statistics for the models 
considered in this paper. The addition of extra free parameters does
of course lead to an increase in the likelihood of the description 
of the data by the model. The message conveyed by the value of the {\tt AIC} 
statistic is less clear. All models show a slight 
decrease in the value of the {\tt AIC} statistic  
compared with the basic-five set, but the {\tt BIC} statistic 
paints a quite different 
picture. Liddle reports that a difference in the {\tt BIC} of 2 should be 
regarded as `positive evidence' and of 6 as `strong evidence' in favour of 
the model with the smaller value of {\tt BIC}. 
Therefore, there is apparently `positive evidence' that we should not expand the 
basic-five model to allow the scalar spectral index to float 
and `strong evidence' that we really should not have burdened the 
reader with the basic-six plus one more free parameter models. 

This conclusion, and indeed the basic {\tt BIC} formula itself, appears
to disagree with the approach of this paper. For data with Gaussian errors,
the addition of a further parameter would be expected to reduce $-2\ln{\cal L}$
by one via the usual `degree of freedom' rule (although note that this
strictly applies only to parameters linearly related to the data, such
as polynomial expansion coefficients). Furthermore, the reduction in 
$-2\ln{\cal L}$ should be distributed as $\chi^2$ with 1 degree of freedom 
if the new parameter is not part of the true model, independent of the 
number of data points. A reduction in $-2\ln{\cal L}$ of 4 therefore 
amounts to rejection at 95\% confidence of the hypothesis that a new 
parameter is not required. Thus, the fact that allowing deviations 
from scale invariance reduces $-2\ln{\cal L}$ by 3.7 amounts to 
marginal evidence for the reality of tilt. This reasoning matches 
the {\tt AIC} approach quite well, as long as the coefficient 2 in 
the $2N_{\rm par}$ term is regarded as being adjustable according to 
the significance level of interest.

The {\tt BIC} statistic is an approximate form of the `Bayesian evidence'
(Hobson, Bridle \& Lahav 2002, Liddle 2004, Trotta 2005). 
One of the conditions that must be satisfied in order for the {\tt BIC} 
to be a good approximation to the Bayesian evidence is the independence 
of the data points under consideration. 
By setting $N_{\rm data}=1403$ in the definition of the {\tt BIC} in 
Eq.~\ref{eq:bic}, we are effectively treating all of the data points 
used in our analysis as independent. Our calculation of the {\tt BIC} 
therefore gives an overly pessimistic impression of the impact of adding  
of new parameters. If, for example, an eigenmode analysis or `radical' 
data compression technique was applied to the full set of CMB data points 
used in our analysis, this would produce a much smaller set of genuinely 
independent data points which fully describe the CMB measurements 
(Bond, Jaffe \& Knox 2000). The values of the {\tt AIC} and {\tt BIC} 
statistics would become closer if recomputed for this `reduced' set of 
datapoints. One might argue that one should compute the Bayesian evidence 
rather than approximations such as the {\tt BIC}. There are two reasons why 
we have not done this. Firstly, the Bayesian evidence is hard to compute 
accurately using MCMC techniques, although fast algorithms are under 
development (Mukherjee, Parkinson \& Liddle 2005). Secondly, the definition 
of the prior on a parameter is part of the model tested in the Bayesian 
evidence approach and we believe that this is a weak point in the method 
for the following reason. Since the choice of prior is arbitrary to some 
extent, it is possible in principle to select a prior such that the 
Bayesian evidence increases upon the addition of new parameters. 
We prefer the effectively frequentist argument of simply requiring a 
reduction in $-2\ln{\cal L}$ of order unity in order to claim the 
detection of another degree of freedom.

\subsection{Details of the evidence for tilt}

\begin{figure}
\includegraphics[width=0.47\textwidth]{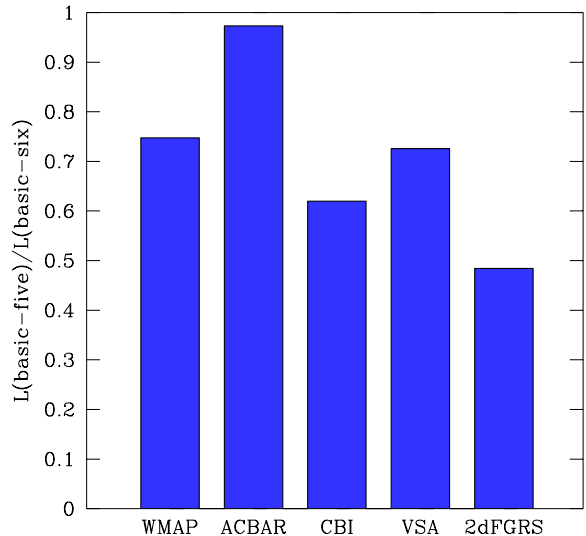}
\caption{
The likelihood ratios of the basic-five model to the basic-six model 
plotted in Fig.\ref{fig:experiments} for the individual datasets used 
in our analysis.
}
\label{fig:b6_split}
\end{figure}

Putting aside the caveat raised by the {\tt BIC} statistic, it is important 
to look at the basic-five and basic-six model results for the scalar spectral 
index in more detail, as they have important implications for inflationary 
models. To recap, in Section~3.2 we set $n_{\rm s}=1$, i.e. the scale invariant 
value of the spectral index for primordial scalar fluctuations. In Section~3.3, 
we treated the spectral index as a free parameter and found that 
$n_{\rm s}=1$ was on the $95\%$ limit. 
Fig.~\ref{fig:experiments} shows the best fitting models to the CMB 
temperature power spectrum data (upper panels) and the 2dFGRS $P(k)$ 
(lower panels) for the basic-five (solid lines) and basic-six (dashed lines) 
models. The difference between the two models is small and comes mostly from
scales beyond those probed by WMAP; this is quantified in the 
right-hand panels in which the models and datapoints have been divided by 
the $n_{\rm s}=1$ model. Fig.~\ref{fig:b6_split} shows the 
likelihood quotients between the basic-five and basic-six models 
for each dataset separately. 
It is clear that the datasets primarily responsible for driving 
the scalar spectral index away from the scale invariant value 
are the CBI measurements and the 2dFGRS $P(k)$:
the basic-six model represents only a modest improvement over the basic-five 
model in its description of the WMAP and VSA datasets, while the addition 
of an extra parameter makes very little difference to how well the ACBAR results 
are reproduced. Nevertheless, there is an impressive consistency between
the various datasets: a systematic error in a single one of these might
have resulted in an improved overall likelihood on the introduction of tilt, but
at the price of a poorer fit to some of the correct datasets. This is
not what we see: addition of the 2dFGRS strengthens a weak trend already
present in the CMB data. Even so, the overall result remains tantalisingly
placed in terms of its statistical significance: 95\% confidence is
not sufficient to claim firm detection of an effect of this importance.
The best that can be said is that even a modest amount of extra data 
could easily move things into the territory of firm detection. The
largest predicted deviations from $n_{\rm s}=1$ occur around the third
CMB peak, at $\ell\simeq 800$, and the data here may be expected to improve
rapidly.

\section{Comparison with constraints obtained using the CMB data and the 
SDSS power spectrum}
\label{sec:sdss}

\begin{figure*}
\includegraphics[width=0.95\textwidth]{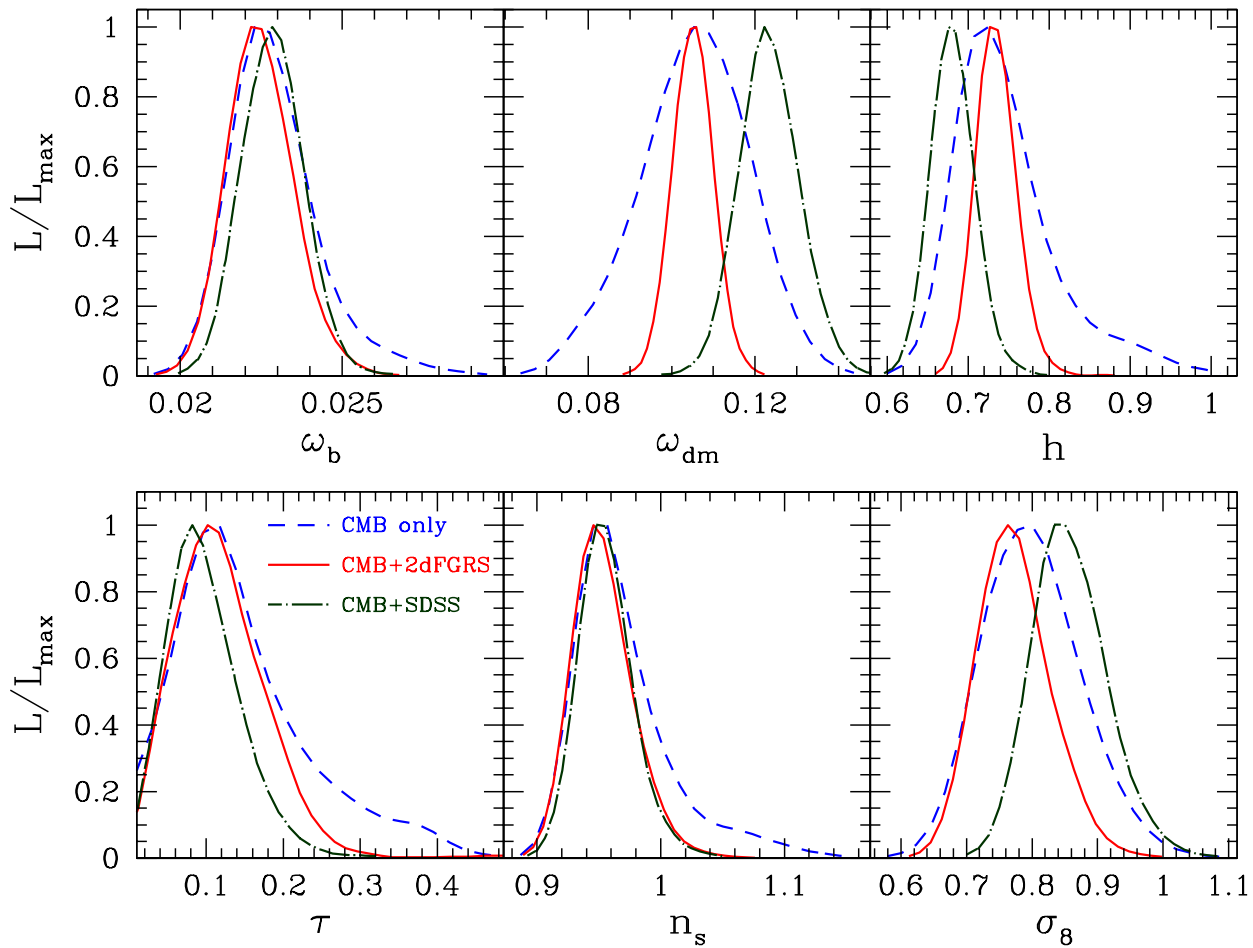}
\caption{
The marginalized one-dimensional posterior likelihood in the basic-six 
parameter space obtained for CMB information only (dashed lines), 
CMB plus the 2dFGRS $P(k)$ (solid lines) and CMB plus the SDSS 
(dot-dashed lines).
}
\label{fig:b6_sdss}
\end{figure*}

\begin{table*}
\renewcommand\arraystretch{1.45}
\begin{center}
\caption{The marginalized 68\% interval constraints (unless otherwise stated) on cosmological parameters 
obtained using CMB data and the SDSS galaxy power spectrum for different 
parameter sets.}
\end{center}
\begin{center}
\begin{tabular}{lcccccc}
\hline\hline
& b5 & b6 & b6 + $f_{\nu}$  & b6 + $\Omega_{k}$ & b6 + $w_{\rm DE}$ & b6 + $r$ \\ 
\hline\hline
$\Omega_{k}$        & 0                            & 0                             & 0                             &
                       $-0.070^{+0.037}_{-0.039}$   & 0                             & 0                              \\ 
$\Theta$             & $1.0493^{+0.0036}_{-0.0036}$ & $1.0436^{+0.0048}_{-0.0047}$  & $1.0436^{+0.0047}_{-0.0048}$  &
                       $1.0408^{+0.0051}_{-0.0052}$ & $1.0440^{+0.0048}_{-0.0049}$  & $1.0452^{+0.0047}_{-0.0047}$ \\ 
$\omega_{dm}$       & $0.1227^{+0.0074}_{-0.0073}$ & $0.1234^{+0.0070}_{-0.0071}$  & $0.1304^{+0.0094}_{-0.0094}$  &
                       $0.106^{+0.010}_{-0.010}$    & $0.104^{+0.012}_{-0.012}$     & $0.1203^{+0.0069}_{-0.0069}$  \\ 
$\omega_{b}$        & $0.0242^{+0.0006}_{-0.0006}$ & $0.0228^{+0.0009}_{-0.0009}$  & $0.0225^{+0.0009}_{-0.0009}$  &
                       $0.0224^{+0.0011}_{-0.0012}$ & $0.0235^{+0.0013}_{-0.0013}$  & $0.0234^{+0.0010}_{-0.0010}$ \\ 
$f_{\nu }$           & 0                            & 0                             & $<0.104$ (95\%)     &
                       0                            & 0                             & 0 \\ 
$\tau $              & $0.173^{+0.034}_{-0.036}$    & $0.097^{+0.046}_{-0.045}$     & $0.098^{+0.043}_{-0.044}$     &
                       $0.147^{+0.071}_{-0.077}$    & $0.124^{+0.065}_{-0.064}$     & $0.099^{+0.046}_{-0.046}$ \\ 
$w_{\rm DE}$             & $-1$                          & $-1$                           & $-1$                           &
                      $-1$                          & $-0.45^{+0.23}_{-0.23}$       & $-1$\\ 
$n_{s}$              & 1                            & $0.956^{+0.020}_{-0.020}$     & $0.947^{+0.022}_{-0.022}$     &
                       $0.958^{+0.026}_{-0.032}$    & $0.988^{+0.040}_{-0.039}$     & $0.974^{+0.024}_{-0.025}$ \\ 
$\log_{10}(10^{10}A_{s})$ & $3.273^{+0.057}_{-0.061}$    & $3.100^{+0.098}_{-0.098}$     & $3.097^{+0.094}_{-0.095}$        &
                       $3.12^{+0.14}_{-0.15}$       & $3.10^{+0.13}_{-0.13}$        & $3.100^{+0.097}_{-0.098}$  \\ 
$r$                  & 0                            & 0                             & 0                             &
                       0                            & 0                             & $<0.31$ (95\%)                   \\ 
\hline
$\Omega_{\rm DE}$       & $0.710^{+0.030}_{-0.030}$    & $0.682^{+0.035}_{-0.035}$     & $0.603^{+0.070}_{-0.073}$     &
                       $0.577^{+0.083}_{-0.088}$    & $0.557^{+0.085}_{-0.085}$     & $0.706^{+0.034}_{-0.034}$    \\
$t_{0}/{\rm Gyr}$    & $13.35^{+0.12}_{-0.12}$      & $13.65^{+0.20}_{-0.20}$       & $13.97^{+0.26}_{-0.27}$       & 
                       $16.2^{+1.2}_{-1.1}$         & $14.35^{+0.56}_{-0.50}$       & $13.54^{+0.21}_{-0.21}$   \\
$\Omega_{\rm m}$           & $0.289^{+0.030}_{-0.030}$    & $0.317^{+0.035}_{-0.035}$     & $0.397^{+0.073}_{-0.070}$     &
                       $ 0.49^{+0.12}_{-0.11}$      & $0.443^{+0.084}_{-0.085}$     & $0.294^{+0.034}_{-0.034}$   \\
$\sigma_8$           & $0.947^{+0.039}_{-0.040}$    & $0.858^{+0.054}_{-0.054}$     & $0.732^{+0.084}_{-0.083}$     &
                       $0.773^{+0.071}_{-0.071}$    & $0.57^{+0.15}_{-0.16}$        & $0.853^{+0.055}_{-0.055}$   \\
$z_{re}$             & $17.5^{+2.3}_{-2.4}$         & $11.9^{+3.9}_{-3.9}$          & $12.2^{+3.8}_{-3.9}$          &
                       $14.8^{+7.1}_{-7.1}$         & $13.4^{+4.7}_{-4.9}$          & $11.8^{+3.8}_{-3.9}$   \\
$h$                  & $0.714^{+0.021}_{-0.021}$    & $0.681^{+0.026}_{-0.026}$     & $0.626^{+0.040}_{-0.042}$     &
                       $0.519^{+0.064}_{-0.067}$    & $0.544^{+0.047}_{-0.047}$     & $0.701^{+0.028}_{-0.029}$   \\
 $\sum{m_{\nu}}/\rm{eV}$ & 0                        & 0
& $<1.27$ (95\%)    &
                       0                            & 0                            & 0                          \\
\hline\hline
\end{tabular}
\end{center}
\end{table*}

In this section, we replace the 2dFGRS $P(k)$ measured by Cole et~al. (2005) 
with the power spectrum of SDSS galaxies estimated by Tegmark et~al. (2004a) 
and examine the impact that this change has upon the values of the recovered 
cosmological parameters. 
There are a number of differences between the two measurements of the galaxy 
power spectrum. Firstly, the SDSS is a red-selected survey, while 
the 2dFGRS is blue-selected. Secondly, Tegmark et~al. 
used a sophisticated eigenmode deconvolution apparatus to attempt
to remove the effects of the survey window and redshift-space distortions;
in contrast, Cole et~al. used a simpler Fourier approach that compares
to window-convolved models and quantifies redshift-space effects directly
by comparison with realistic simulations.

Tegmark et~al. (2004b) used the WMAP first year data and the SDSS galaxy 
power spectrum to constrain cosmological parameters. 
These authors modelled the galaxy power spectrum with a non-linear model 
for the matter fluctuations multiplied by a scale independent bias factor. 
The power spectrum data were used on scales larger than 
$k<0.20 \,h\, \mathrm{{Mpc^{-1}}}$. 
It is not clear that a constant bias is a good approximation on scales 
for which the density fluctuations have become non-linear. 
We adopt a simpler approach and assume that the galaxy power spectrum can 
be related to the linear perturbation theory power spectrum of the mass 
through a constant shift in amplitude. 
As discussed earlier, the simulations used by Cole et~al. 
indicate that  redshift-space effects and other nonlinearities
are unimportant for the 2dFGRS to our imposed limit of $k_{\rm max}=0.15\, h\,{\rm Mpc}^{-1}$.

We repeat the study of parameter space previously carried out using the 
CMB data plus the 2dFGRS $P(k)$ and present our 
results using the SDSS $P(k)$ instead in Table~5. 
For the most part, the results obtained with the SDSS $P(k)$ are compatible 
with those found using the 2dFGRS $P(k)$. There are, however, some cases in 
which the results are quite different. This point is illustrated using 
the results of the basic-six model in Fig.~\ref{fig:b6_sdss}. In this plot, 
we compare the parameter constraints obtained using CMB data alone (dashed 
line) with the results for CMB data plus the 2dFGRS $P(k)$ (solid line) and 
for CMB data plus the SDSS $P(k)$ (dot-dashed line). 
For three out of the six parameters presented, $\omega_{\rm b}$, $\tau$ and 
$n_{\rm s}$, there is impressive agreement between the sets of results; the 
peaks in the likelihood distributions coincide in the CMB only and CMB plus 
$P(k)$ cases, and the results are consistent to better than the 68\% 
confidence intervals. 
The agreement between the sets of results for the scalar spectral index 
in particular is excellent. However, for the other three parameters plotted, 
$\omega_{\rm dm}$, $h$ and $\sigma_{8}$, the agreement is less impressive; 
the differences in the recovered values of  $h$ and $\sigma_{8}$ are driven 
by the change in $\omega_{\rm dm}$.
The peak in the likelihood distributions for the CMB only and CMB plus 
2dFGRS $P(k)$ cases are in good agreement, as remarked upon in 
Section~3. There is a clear discrepancy, however, with the preferred parameter 
values when using CMB data plus the SDSS $P(k)$. This is most marked for 
the physical density of dark matter, $\omega_{\rm dm}$. Cole et~al. (2005) 
noted that the SDSS $P(k)$ has a slightly bluer slope than that
of the 2dFGRS, favouring higher values of $\Omega_{\rm m}$. 

There are two other notable discrepancies between the results obtained 
with the SDSS and 2dFGRS $P(k)$ in our basic-six plus one additional 
free parameter models. When the assumption of a flat universe is 
relaxed, we find that the constraints on $\Omega_{k}$ are weaker 
in the SDSS case, $\Omega_{k}=-0.070_{-0.039(0.079)}^{+0.037(0.058)}$; 
the allowed range is nearly twice as broad as in the case of the 
2dFGRS $P(k)$. This is because on the scales used in our analysis, 
the SDSS power spectrum does a poorer job of constraining $\Omega_{\rm m}$ 
compared with the 2dFGRS $P(k)$, and hence is not as effective at breaking the 
geometrical degeneracy.
In the basic-six plus dark energy equation of state parameter set, we find 
$w_{\rm DE}=-0.45_{-0.23}^{+0.23}$ using the SDSS $P(k)$, much higher than we 
found in the case of the 2dFGRS $P(k)$ and inconsistent with a cosmological 
constant. 
If we also include the SNIa data from Riess et~al. (2004), then 
we obtain a value for the equation of state that is consistent with our 
previous results: $w_{\rm DE}=-0.89_{-0.18}^{+0.19}$. Again, the discrepancy 
in the result for the equation of state can be traced back to the preferred 
values of $\omega_{\rm dm}$. 
We note that MacTavish et~al. (2005) find similar results to ours 
for the equation of state of the dark energy using the SDSS galaxy 
power spectrum. 
Fig. 8 shows the degeneracy in the $\Omega_{\rm m} - w_{\rm DE}$ plane 
for CMB data alone. Adding information from the galaxy power spectrum breaks 
this degeneracy. If the galaxy $P(k)$ data prefer a high value of 
$\Omega_{\rm m}$, as is the case for the SDSS data, then a high value of 
$w_{\rm DE}$ will result. 

It may be that these differences between 2dFGRS and SDSS amount to no
more than an unlucky amount of cosmic variance, but clearly it would
be more reassuring if the results showed greater consistency. It will
therefore be important to see each dataset subjected to analysis
by a variety of algorithms and codes, as happened following the
Percival et~al. (2001) 2dFGRS power-spectrum analysis
(Tegmark, Zaldarriaga \& Hamilton 2001). This older comparison
found consistent results, but the comparison will now be more
demanding, given the smaller errors arising from
current datasets.

\section{Summary}

We have placed new constraints on the values of the basic cosmological 
parameters, using an up-to-date compilation of CMB data and the galaxy 
power spectrum measured from the final 2dFGRS by Cole et~al. (2005). 
We have carried out a comprehensive exploration of parameter space, 
considering five, six and seven parameter models, making different 
assumptions about the priors used for certain parameters. 

Our main results can be summarized as follows: 
\begin{itemize} 
\item A model in which five parameters are allowed to vary does a 
remarkably good job of describing the currently available CMB and 
LSS data. 
\item There is an impressive level of agreement between the results 
obtained for CMB data alone and for CMB data plus the 2dFGRS power 
spectrum data. If the 2dFGRS $P(k)$ is replaced by the SDSS $P(k)$, 
there is some tension between the parameter values preferred by 
the CMB and SDSS datasets. 
\item For some parameters, for example the physical density of dark 
matter, Hubble's constant and the amplitude of density fluctuations, 
there is a significant tightening of the allowed range of parameter space 
when the 2dFGRS $P(k)$ is included in the analysis. In particular,
we infer a density significantly below $\Omega_{\rm m}=0.3$.  
\item We find some evidence for a departure from a scale invariant 
primordial spectrum of scalar fluctuations. Our results for the scalar 
perturbation spectral index are only marginally consistent with the 
scale invariant value $n_{\rm s}=1$; this spectrum is formally excluded 
at the $95\%$ confidence level.
However, this conclusion is weakened if we drop the assumption 
that the universe is flat or allow neutrinos to have a mass.  
\item We place new limits on the mass fraction of massive neutrinos: 
$f_{\nu} < 0.105$ and $\sum m_{\nu} < 1.2$~eV at the 95\% level. 
\item Several parameters are sensitive to the choice of prior for 
the optical depth to the last scattering surface, $\tau$. 
\item We find that a wide range of closed universes are consistent 
with the CMB data. This range is restricted if we also consider 
the 2dFGRS $P(k)$ data. If we further assume a prior of $\tau < 0.3$, 
then the preferred spatial curvature is close to flat.  
\item We confirm the evidence previously reported by Efstathiou et~al. (2002) 
for a non-zero dark energy contribution to the energy-density of the 
universe. 
\item We find a redshift-independent equation of state for the dark 
energy of $w_{\rm DE} = -0.85_{-0.17}^{+0.18}$, consistent with 
a cosmological constant. 
\item Inflationary models with a scalar field potential with a 
$V(\phi) \propto \phi^{4}$ term are  
ruled out by our analysis. 
\end{itemize} 

The final message of this analysis is that 
meaningful comparison of the parameter constraints from
different studies requires a clear listing of the free
parameters and their prior distributions. Although current
datasets measure many parameter combinations extremely well,
important degeneracies remain. As we have discussed, there is
a good chance that current measurements may be poised on
the brink of rejecting the simplest 5-parameter model in
favour of something more complicated. However, even if this
step is taken, it will require much work before any such
deviation from the standard model could have a unique interpretation.

\section*{Acknowledgements}

We would like to thank Sarah Bridle for her kind and invaluable help 
and useful discussions and the referee, Andrew Jaffe, for a constructive 
report. We also aknowledge Andrew Liddle for discussions about Bayesian 
Evidence. AGS acknowledges the hospitality of the Department of Physics 
at the University of Durham where part of 
this work was carried out; he would also like to thank the scientists  
throughout the world who submit their articles to the arXiv preprint 
data base, thereby making them freely available to scientists in 
developing countries. 
AGS acknowledges a fellowship from CONICET; 
CMB is funded by a Royal Society University Research Fellowship; 
JAP holds a PPARC Senior Research Fellowship;
WJP acknowledges receipt of a PPARC fellowship;
NDP is funded by Fondecyt project number 3040038;
PN acknowledges receipt of a ETH Zwicky Fellowship.
This work was supported by the European Commission's ALFA-II programme 
through its funding of the Latin-american European Network for Astrophysics 
and Cosmology, LENAC, and by PPARC.





\def\japref{\parskip=0pt\par\noindent\hangindent\parindent}

\section*{References}

\japref  Abazajian K. et~al., 2005, AJ, 129, 1755

\japref  Ahmad Q.R., Allen R.C., Andersen T.C., 2001,  Phys. Rev. Letters, 
87, 071301

\japref  Allen S.W., Schmidt R.W., Bridle S.L., 2003, MNRAS, 346, 593

\japref  Akaike H., 1074, IEEE Trans. Auto. Control, 19, 716

\japref  Bacon D.J., Massey R.J., Refregier A.R., Ellis R.S., 2003, 
         MNRAS, 344, 673

\japref  Barger V., Marfatia D.,  Whisnant K., 2003, Int. J. Mod. Phys., 
E12, 569

\japref  Bennett C.L., et~al. , 2003, ApJS, 148, 1

\japref  de Bernardis P., et~al., 2000, Nature, 
404, 955 


\japref  Bond J.R., Jaffe A.H., Knox, L., 2000, ApJ, 533, 19

\japref  Bridle S.L., Lahav O., Ostriker J.P., Steinhardt P.J., 2003a, 
Science, 299, 1532

\japref  Bridle S.L., Lewis A.M., Weller J., Efstathiou G., 2003b, MNRAS, 
         342, L72. 

\japref  Cole S., et~al. 2005, MNRAS, 362, 505

\japref  Colless M., et~al., 2001, MNRAS, 328, 1039

\japref  Colless M., et~al., 2003, astro-ph/0306581

\japref  Cuoco A., Iocco F., Magnano G., Miele G., Pisanti O., Serpico
P.D., 2004, Int. J. Mod. Phys., A19, 4431

\japref  Croft R.A.C., Weinberg D.H., Bolte M., Burles S., Hernquist L. 
Katz N., Kirkman D., Tytler D., 2002, ApJ, 581, 20

\japref  de Oliveira-Costa A., Tegmark M., Zaldarriaga M., Hamilton A., 
         2004, Phys. Rev. D., 69, 063516


\japref  Dickinson C., et~al. , 2004, MNRAS, 353, 732

\japref  Efstathiou G., 2003, MNRAS, 343, L95 

\japref  Efstathiou G., 2004, MNRAS, 348, 885

\japref  Efstathiou G., Bond J.R., 1999, MNRAS, 304, 75

\japref  Efstathiou G., et~al. 2002, MNRAS, 330, L29

\japref  Elgaroy O., et~al., 2002, Phys. Rev. Letters, 89, 061301

\japref  Freedman W.L., et~al., 2001, ApJ, 553, 47

\japref  Frenk C.S., White S.D.M., Davis M., 1983, ApJ, 271, 417

\japref  Gazta\~{n}aga E., Wagg J., Multamaki T., Montana A., Hughes D.H., 
         2003, MNRAS, 346, 47

\japref  Gazta\~{n}aga E., Norberg P., Baugh C.M., Croton D.J., 2005, 
         MNRAS, 364, 620

\japref  Gelman A., Rubin D., 1992, Statistical Science, 7, 457

\japref  Gnedin N.Y., Hamilton A.J.S., 2002, MNRAS, 334, 107. 

\japref  Hanany S., et~al., 2000, 545, L5.

\japref  Hannestad S., 2002, Phys. Rev. D., 66, 125011

\japref  Hansen F.K., Balbi A., Banday A.J., Gorski K.M., 2004, MNRAS, 
354, 905

\japref  Henry J.P., 2004, ApJ, 609, 603

\japref  Heymans C. et~al., 2005, MNRAS, 361, 160 

\japref  Hinshaw G., et~al., 2003, ApJS, 148, 135

\japref  Hobson M.P., Bridle S.L., Lahav O., 2002, MNRAS, 335, 377

\japref  Hoekstra H., Yee H.K., Gladders M.D., 2002, ApJ, 577, 595 

\japref  Hu W., Eisenstein D.J., Tegmark M., 1998, Phys. Rev. Letters, 80, 5255

\japref  Kogut A., et~al., 2003, ApJS, 148, 161

\japref  Komatsu E., et~al. 2003, ApJS, 148, 119

\japref  Kosowsky A., Milosavljevic M., Jimenez R., 2002, Phys. Rev. D., 66, 063007

\japref  Kuo C.L., et~al., 2004, ApJ, 600, 32

\japref  Leach S.M., Liddle A.R., 2003a, MNRAS, 341, 1151

\japref  Leach S.M., Liddle A.R., 2003b, Phys. Rev. D., 68, 123508

\japref  Lewis A., Bridle S., 2002, Phys. Rev. D, 66, 103511

\japref  Lewis A., Challinor A., Lasenby A., 2000, Apj 538, 473

\japref  Liddle A.R., 2004, MNRAS, 351, L49

\japref  MacTavish C.J. et~al., 2005, ApJ submitted, astro-ph/0507503

\japref  McDonald P., Seljak U., Cen R., Bode P., Ostriker J.P., 2005, 
            MNRAS, 360, 147. 

\japref  Metropolis N., Rosenbluth A., Rosenbluth R., Teller A., Teller E., 
 1953, J. Chem. Phys., 21, 1087

\japref  Mukhanov V.J., Feldman H.A., Brandenberger R.H., 1992, Phys.
Rep., 215, 203 

\japref  Mukherjee P, Parkinson D, Liddle A.R., 2005, MNRAS, submitted.
(astro-ph/0508461)

\japref  Percival W.J., et~al., 2001, MNRAS, 327, 1297

\japref  Percival W.J., et~al., 2002, MNRAS, 337, 1068

\japref  Percival W.J., et~al., 2004, MNRAS, 353, 1201

\japref  Perlmutter S., et~al., 1999, ApJ, 517, 565

\japref  Peiris H.V., et~al., 2003, ApJS, 148, 213

\japref  Pope A.C. et al.,  ApJ, 607, 655 

\japref  Readhead A.C.S. et~al., 2004, ApJ, 609, 498

\japref  Riess A.G., et~al., 2004, ApJ, 607, 665

\japref  Sahni V., 2005, in Papantonopoulos E. ed., The Physics of the Early
Universe. Springer, Berlin, p. 141 

\japref  Schwarz D. J., Terrero-Escalante C. A., Garcia A. A., 
         2001, Phys. Let. B., 517, 243

\japref  Schwartz G., 1978, Annals of Statistics, 5, 461

\japref  Seljak U. et~al., 2005, Phys. Rev. D, 71, 103515

\japref  Slosar A., Seljak U., Makarov A., 2003, Phys. Rev. D., 69, 123003

\japref  Spergel D.N., et~al., 2003, ApJS, 148, 175

\japref  Tegmark M., Zaldarriaga M., Hamilton A.J.S., 2001, MNRAS,335, 887

\japref  Tegmark M. et~al., 2004a, ApJ 606, 702

\japref  Tegmark M. et~al., 2004b, Phys. Rev. D, 69,  103501

\japref  Trotta R, 2005, MNRAS, submitted. (astro-ph/0504022)

\japref  Verde L., et~al., 2002, MNRAS, 335, 432

\japref  Verde L., et~al., 2003, ApJS, 148, 195 

\japref  Weinheimer C., 2003, Nucl. Phys. B (Proc. Suppl.), 118, 279


\japref  York D., et~al., 2000, AJ, 120, 1579


\end{document}